\documentclass[11pt,a4paper]{article}
\pdfoutput=1

\usepackage{graphicx} 
\usepackage{mathtools} 
\usepackage{amssymb}
\usepackage{bm}
\usepackage{accents}
\usepackage{color}
\usepackage[bookmarks=true,hyperfigures=true]{hyperref}
\usepackage[nosort]{cite}
\usepackage[bulletsep]{collref}
\usepackage{tensor}
\usepackage[english]{babel}
\usepackage[section]{placeins}

\usepackage[a4paper,text={165mm,225mm},centering]{geometry}

\allowdisplaybreaks[3]

\numberwithin{equation}{section}

\newcommand{\nn}{\nonumber}

\usepackage[font=small,labelfont=bf,width=0.85\textwidth]{caption}

\let\oldbfseries=\bfseries
\renewcommand{\bfseries}{\oldbfseries\boldmath}

\newcommand{\sfrac}[2]{{\textstyle\frac{#1}{#2}}}
\newcommand{\half}{\sfrac{1}{2}}

\DeclareMathOperator{\tr}{tr}

\newcommand*{\diff}{{\mathrm d}}

\newcommand{\ft}[2]{{\textstyle\frac{#1}{#2}}}
\newcommand*{\pexp}{{\mathcal{P} \exp}}

\newcommand{\one}{\vert 1 \rangle}
\newcommand{\two}{\vert 2 \rangle}
\newcommand{\zb}{\bar{z}}
\newcommand{\LO}{\mathrm{O}}
\newcommand{\Li}{\mathrm{Li}}
\newcommand{\Img}{\mathrm{Im}}

\newcommand{\Gacr}{\Gamma_{\mathrm{cross}}}
\newcommand{\Gacu}{\Gamma_{\mathrm{cusp}}}
\newcommand{\gacu}{\gamma_{\mathrm{c}}}
\newcommand{\gacur}{\gamma_{\mathrm{c,r}}}

\newcommand{\Gacrt}{\widetilde{\Gamma}_{\mathrm{cross}}}
\newcommand{\Gacrh}{\widehat{\Gamma}_{\mathrm{cross}}}
\newcommand{\Ft}{\widetilde{F}}
\newcommand{\Ct}{\widetilde{C}}

\begin{document}
\thispagestyle{empty}

\begingroup\raggedleft\footnotesize\ttfamily
HU-EP-18/15\
\vspace{15mm}
\endgroup

\begin{center}
{\Large\bfseries The Cross Anomalous Dimension in Maximally Supersymmetric Yang--Mills Theory\par}
\vspace{25mm}

\begingroup\scshape\large 
Hagen M\"unkler
\endgroup
\vspace{5mm}

\textit{Institut f\"ur Theoretische Physik, Eidgen\"ossische Technische Hochschule Z\"urich, \phantom{$^\S$}\\
Wolfgang-Pauli-Strasse 27, CH-8093 Z\"urich, Switzerland. \phantom{$^\S$}\\
and \phantom{$^\S$}\\
Institut f\"ur Physik and IRIS Adlershof, Humboldt--Universit\"at zu Berlin, \phantom{$^\S$}\\
Zum Gro{\ss}en Windkanal 6, D-12489 Berlin, Germany, 
} \\[0.1cm]
\texttt{ \small{ muenkler@itp.phys.ethz.ch \phantom{\ldots}}} \\ \vspace{5mm}

\vspace{18mm}

\textbf{Abstract}\vspace{5mm}\par
\begin{minipage}{14.7cm}
The cross or soft anomalous dimension matrix describes the renormalization of Wilson loops with a self-intersection
and is an important object in the study of infrared divergences of scattering amplitudes. In this 
paper it is studied for the Maldacena--Wilson loop in $\mathcal{N} \! =4$ supersymmetric Yang--Mills theory and 
Euclidean kinematics. We consider both the strong-coupling description in terms of minimal surfaces in $\mathrm{AdS}_5$ 
as well as the weak-coupling side up to the two-loop level. In either case, the coefficients of the cross anomalous 
dimension matrix can be expressed in terms of the cusp anomalous dimension. The strong-coupling description displays 
a Gross--Ooguri phase transition and we argue that the cross anomalous dimension is an interesting object to study in an 
integrability-based approach.
\end{minipage}\par
\end{center}
\newpage

\setcounter{tocdepth}{1}
\hrule height 0.75pt
\tableofcontents
\vspace{0.8cm}
\hrule height 0.75pt
\vspace{1cm}



\section{Introduction}

Recent years have witnessed much progress in our understanding of the soft and collinear singularities of gauge theory 
scattering amplitudes. This progress is of phenomenological relevance in QCD, where our control over these singularities 
allows for the resummation of logarithmically enhanced contributions. The key element in the description of the infrared 
singularities is the soft or cross anomalous dimension matrix, which describes the singularities of $n$-leg scattering 
amplitudes via the correlation function of $n$ semi-infinite Wilson lines with a common starting point. 
In the case of massless external particles corresponding to lightlike Wilson lines, there exist an intriguingly 
simple ansatz for the soft anomalous dimension matrix known as the dipole formula 
\cite{Becher:2009cu,Gardi:2009qi,Becher:2009qa}. It fixes the kinematic dependence of the soft anomalous dimension 
in terms of the cusp anomalous dimension. However, starting at the three-loop level, deviations from the dipole 
formula begin to occur \cite{Almelid:2015jia}. In the case of massive external particles corresponding to 
timelike Wilson lines, the soft anomalous dimension contains kinematic dependences which are not related to 
the cusp anomalous dimension already at the two-loop level \cite{Mitov:2009sv}.  

In the present paper, we discuss the cross anomalous dimension%
\footnote{We use the term cross anomalous dimension within this 
paper, emphasizing the Wilson loop perspective.}
in $\mathcal{N} \! =4$ supersymmetric Yang--Mills (SYM) theory for the Maldacena--Wilson loop 
\cite{Maldacena:1998im,Rey:1998ik}. In Euclidean signature, it is given by \cite{Drukker:1999zq}
\begin{align}
W ( C ) = \frac{1}{N} \tr \left[ 
	\pexp {\left( i \int _C \diff \tau 
	\left( A_\mu \dot{x}^\mu + i \Phi_I n^I \lvert \dot{x} \rvert \right) \right)}
	\right] .
\end{align}
Here, the scalar fields $\Phi_I$ couple to a contour in $\mathrm{S}^5$, which is described by the vector $n^I(\tau)$. 
The scalar couplings can be seen to originate from the dimensional reduction of a lightlike Wilson loop in 
ten-dimensional $\mathcal{N} \! =1$ SYM theory. Due to the local supersymmetry following from this construction 
\cite{Zarembo:2002an} and the additional structures present in $\mathcal{N} \! =4$ SYM theory, the results might
display any underlying structure more clearly than their counterparts in QCD. 
Moreover, the cross anomalous dimension is an interesting quantity within $\mathcal{N} \! =4$ SYM theory and 
could allow for interesting tests of the AdS/CFT correspondence, if it allows for calculations based on 
integrability \cite{Beisert:2010jr} or localization \cite{Pestun:2016zxk,Zarembo:2016bbk}. 

The singularities arising due to self-intersections have been studied within $\mathcal{N} \! =4$ SYM theory 
in the case of Wilson loops over light-like polygonal contours \cite{Dorn:2011ec,Dorn:2011gf,Dixon:2016epj}, 
which are of particular interest due to the duality to certain scattering amplitudes 
\cite{Alday:2007hr,Brandhuber:2007yx,Drummond:2008aq}. Here, we begin the study of the cross anomalous dimension 
in Euclidean kinematics. Different signatures can be reached via analytic continuation in the cusp angle. 
Our analysis is limited to the crossing of two straight lines at an intersection angle $\phi$ (as well as an 
angle $\rho$ between the different scalar couplings) and a particular crossing of cusped lines which can also 
be described in terms of two angles $\phi$ and $\rho$. The first of these two cases has been studied in QCD in 
refs. \cite{Korchemsky:1993hr,Korchemskaya:1994qp} up to the two-loop level.  

Let us briefly explain how this paper is structured. We begin by reviewing the renormalization properties of 
self-intersecting Wilson loops in section \ref{sec:Review}. 
The minimal surfaces appearing in the strong-coupling description of the Maldacena--Wilson loop are discussed 
in section \ref{sec:Planar}, where we note that the latter of the two cases we discuss displays a 
Gross--Ooguri phase transition. Combining the minimal surfaces with the known information about the cross anomalous 
dimension in the planar limit \cite{Korchemskaya:1994qp}, we then establish the cross anomalous dimension at 
strong coupling, where it can be expressed in terms of the cusp anomalous dimension. 
The result shows that the Gross--Ooguri transition for the minimal surfaces leads to a first-order phase 
transition for the cross anomalous dimension in the limit of infinite 't Hooft coupling $\lambda$. 

We then turn to the perturbative calculation, discussing the one- and two-loop level in sections \ref{sec:OneLoop} 
and \ref{sec:TwoLoop}, respectively. Also at weak coupling it turns out to be possible to express the cross 
in terms of the cusp anomalous dimension. This is trivial at the one-loop level but requires a number of 
cancellations at the two-loop level and is indeed not the case in QCD \cite{Korchemskaya:1994qp}.   
The technical details of the two-loop calculation are collected in appendices \ref{app:Color} and \ref{app:Kinematic}.

\section{The Cross Anomalous Dimension}
\label{sec:Review}

\begin{figure}[t]
\centering
\includegraphics[width=100mm]{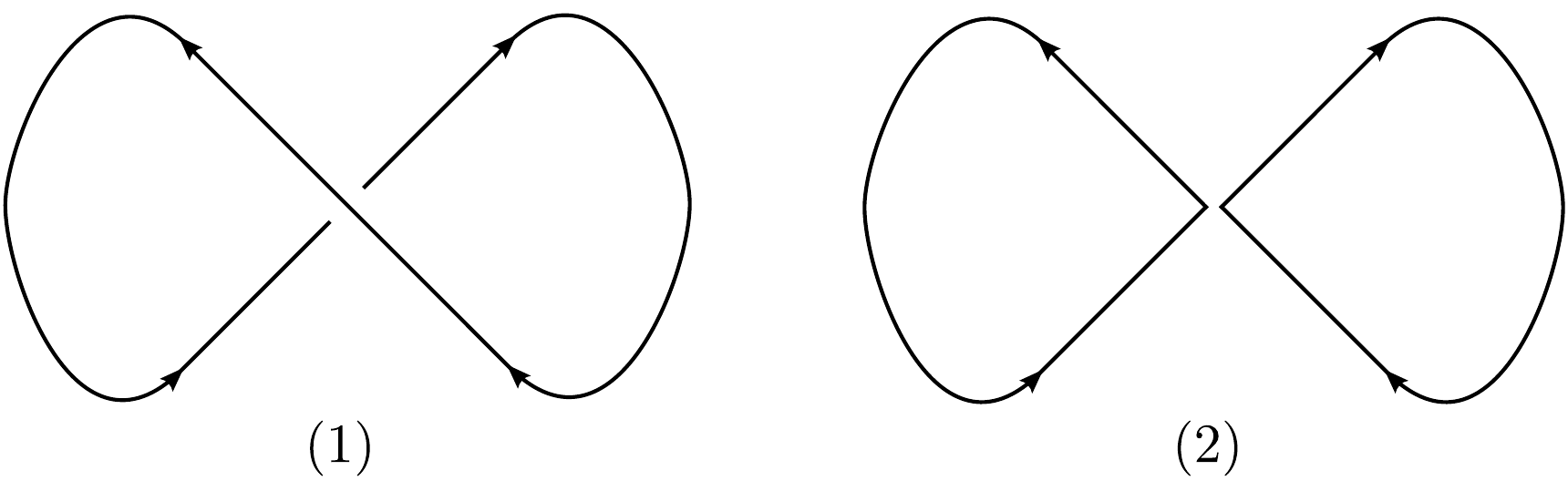}
\caption{The relevant contours for the multiplicative renormalization of self-intersecting Wilson loops.}
\label{fig:mixing}
\end{figure}

\begin{figure}
\centering
\includegraphics[width=80mm]{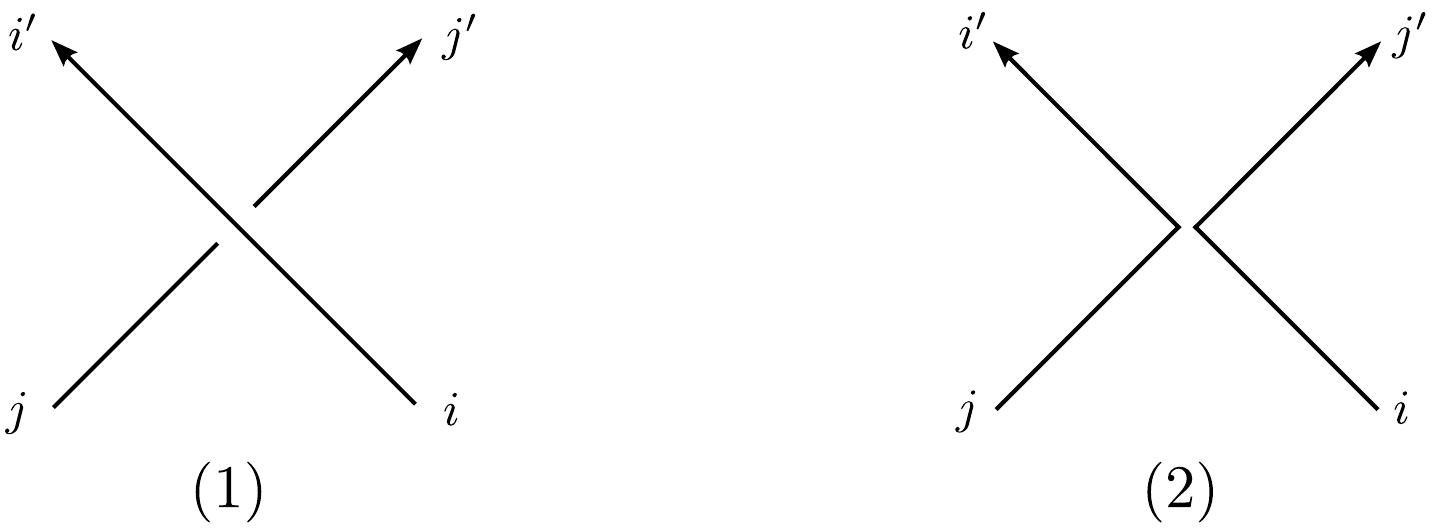}
\caption{Depiction of the contours $C_i$ and the associated color indices.}
\label{fig:mixing2}
\end{figure}

The renormalization of Wilson loops with self-intersections was described in ref.\ \cite{Brandt:1981kf} building on the renormalization for smooth and cusped Wilson loops \cite{Polyakov:1980ca,Dotsenko:1979wb}. It was found that the renormalization requires a mixing between Wilson loops with different path-orderings at the intersection point. In the case of a single intersection, the renormalization mixes the Wilson loop for a self-intersecting eight-shaped contour with the correlator of two Wilson loops, each taken over half of the original contour, cf.\ figure \ref{fig:mixing}. The renormalization hence mixes between the operators
\begin{align}
W_1 &= 	W \left( \raisebox{-1.3mm}{\includegraphics[height=2.8ex]{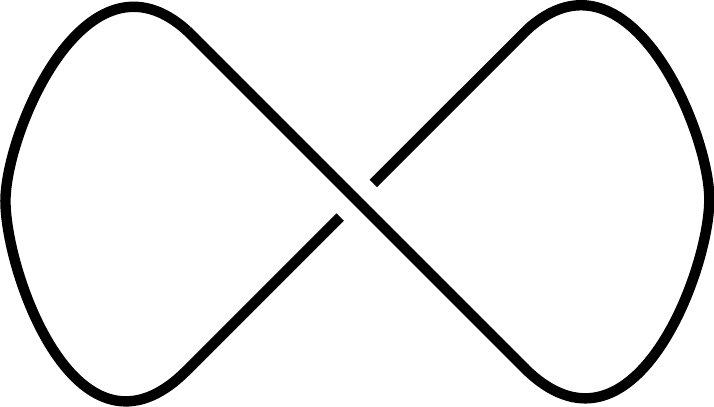}} \right) , & 
W_2 = W \left( \raisebox{-1.3mm}{\includegraphics[height=2.8ex]{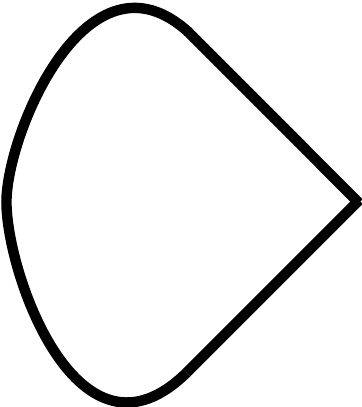}} \right)
	W \left( \raisebox{-1.3mm}{\includegraphics[height=2.8ex]{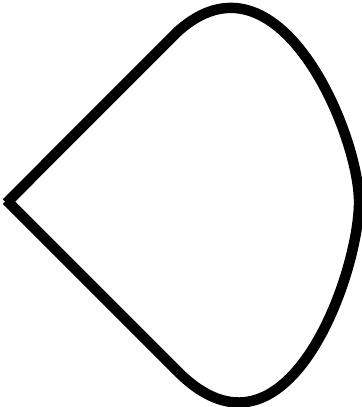}}  \right) .
\end{align} 
The operators $W_1$ and $W_2$ are then renormalized multiplicatively by
\begin{align}
W^R _a = Z_{\mathrm{cross}} {} _{, ab} (\phi, \rho, \lambda) \, W_b \, . 
\end{align}
Here, the $Z$-factor depends on the intersection angle $\phi$ as well as the angle $\rho$ between the $\mathrm{S}^5$-vectors $n_1$ and $n_2$ at the crossing. The cross anomalous dimension $\Gacr$ can be determined from the renormalized Wilson loops by using the renormalization group equation
\begin{align}
\mu \, \frac{\diff}{\diff \mu} W^R _a = \Gacr {} _{, ab} (\phi, \rho, \lambda) \,  W^R _b \, ,
\end{align}
which takes a particularly simple form in $\mathcal{N}=4$ SYM theory due to the vanishing of the beta function. For both the $Z$-factor and the anomalous dimension, the matrix structure is determined by considering the color structure of the diagrams contributing to the expectation value of the operators $W_a$. In order to facilitate the discussion of the color structures of individual diagrams, it is helpful to consider Wilson line operators which are open in color space as it was done in ref.\ \cite{Korchemskaya:1994qp}. We are thus considering the case of two intersecting Wilson lines along $v_1$ and $v_2$ with constant $\mathrm{S}^5$-vectors $n_1$ and $n_2$. The relevant curves are parametrized by
\begin{align}
( x(\sigma) , n(\sigma) )  = \begin{cases}
	(v_1 \sigma \, , \, n_1 ) 
	\quad &\text{for} \quad C_1 \, , \\
	(v_2 \sigma \, , \, n_2 )
	\quad &\text{for} \quad C_2 \, , \\
	(\theta(-\sigma) \, v_1 \sigma  + \theta(\sigma) \, v_2 \sigma \, , \, 
	\theta(-\sigma) \, n_1 + \theta(\sigma) \, n_2 )
	\quad &\text{for} \quad C_3 \, ,\\
	(\theta(-\sigma) \, v_2 \sigma  + \theta(\sigma) \, v_1 \sigma \, , \,
	\theta(-\sigma) \, n_2  + \theta(\sigma) \, n_1  )   
	\quad &\text{for} \quad C_4 \, .
\end{cases}
\end{align}
Here, $v_1$ and $v_2$ are normalized to 1, such that the curves are parametrized by arc-length. The parameter $\sigma$ extends from $-L$ to $L$, such that the lines have finite length. The finite length of the curves serves as an infrared regulator in the calculation of the expectation value, in the case of an infinitely extended line one needs to employ a different regulator such as a fictitious gluon mass. The contours $C_i$ together with the indices of the open Wilson lines in color space are depicted in figure \ref{fig:mixing2}. We are hence considering the expectation values
\begin{align}
\mathcal{W}_1 {} _{i j i^\prime j^\prime} &= \left\langle 
\pexp {\left( i \hspace*{-1mm} \int _{C_1} \hspace*{-2.5mm} \diff \tau 
	\left( A_\mu \dot{x}^\mu + i \Phi_I n^I 
	\lvert \dot{x} \rvert \right) \right)}{} _{j j^\prime} \;
\pexp {\left( i \hspace*{-1mm} \int _{C_2} \hspace*{-2.5mm} \diff \tau 
	\left( A_\mu \dot{x}^\mu + i \Phi_I n^I 
	\lvert \dot{x} \rvert \right) \right)}{} _{i i^\prime}   
\right\rangle , \\
\mathcal{W}_2 {} _{i j i^\prime j^\prime} &= \left\langle 
\pexp {\left( i \hspace*{-1mm} \int _{C_3} \hspace*{-2.5mm} \diff \tau 
	\left( A_\mu \dot{x}^\mu + i \Phi_I n^I 
	\lvert \dot{x} \rvert \right) \right)}{} _{j i^\prime} \;
\pexp {\left( i \hspace*{-1mm} \int _{C_4} \hspace*{-2.5mm} \diff \tau 
	\left( A_\mu \dot{x}^\mu + i \Phi_I n^I 
		\lvert \dot{x} \rvert \right) \right)}{} _{i j^\prime}   
\right\rangle .
\end{align}
Note that while the quantities $\mathcal{W}_1$ and $\mathcal{W}_2$ are not expected to be gauge-invariant, the cross anomalous dimension derived from them is. At the lowest order, the expressions for the functions $\mathcal{W}_1$ and $\mathcal{W}_2$ are trivial, we find the basic color structures
\begin{align}
\mathcal{W}_1  {} _{i j i^\prime j^\prime} 
	= \delta _{i i^\prime} \, \delta _{j j^\prime} =: \left \vert 1 \right \rangle , \qquad 
\mathcal{W}_2 {} _{i j i^\prime j^\prime}  
	= \delta _{i j^\prime} \, \delta _{j i^\prime} =: \left \vert 2 \right \rangle .
\end{align}
The renormalization for $\mathcal{W}_1$ and $\mathcal{W}_2$ is again multiplicative,
\begin{align}
\mathcal{W}^R _a = \widehat{Z}_{\mathrm{cross}} {} _{, ab} (\phi , \rho) \, \mathcal{W}_b \, , 
\end{align}
and the associated anomalous dimension $\Gacrh$ is read off from the renormalization group equation
\begin{align}
\mu \, \frac{\diff}{\diff \mu} \mathcal{W}^R _a = \Gacrh {} _{, ab} (\phi, \rho) \,  \mathcal{W}^R _b \, . 
\label{rge:gacrt}
\end{align}
We can identify the expectation values of the Wilson loop operators $W_1$ and $W_2$ with the open lines by closing the contours in color space, which amounts to contracting them with $\delta _{i j^\prime} \, \delta _{j i^\prime} $, 
\begin{align}
\left \langle W_1 \right \rangle \simeq \frac{1}{N} \langle 2 \vert \mathcal{W}_1 \, , \qquad
\left \langle W_2 \right \rangle \simeq \frac{1}{N^2} \langle 2 \vert \mathcal{W}_2 \, .
\end{align}
While the identification misses finite pieces associated to the closing of the contour, the anomalous dimension is unaffected as it only depends on the angles at the intersection point. We then find that the anomalous dimension matrices $\Gacr$ and $\Gacrh$ are related by a similarity transformation,
\begin{align}
\Gacr = S \, \Gacrh S^{-1} \, \qquad \text{where} \qquad
S = \begin{pmatrix}
\sqrt{N} & 0 \\
0 & 1/ \sqrt{N}
\end{pmatrix} .
\label{rel:gacr}
\end{align}
The $Z$-factors are related in the same way.

\section{The Planar Limit and Strong Coupling}
\label{sec:Planar}

\begin{figure}[t]
\centering
\includegraphics[width=40mm]{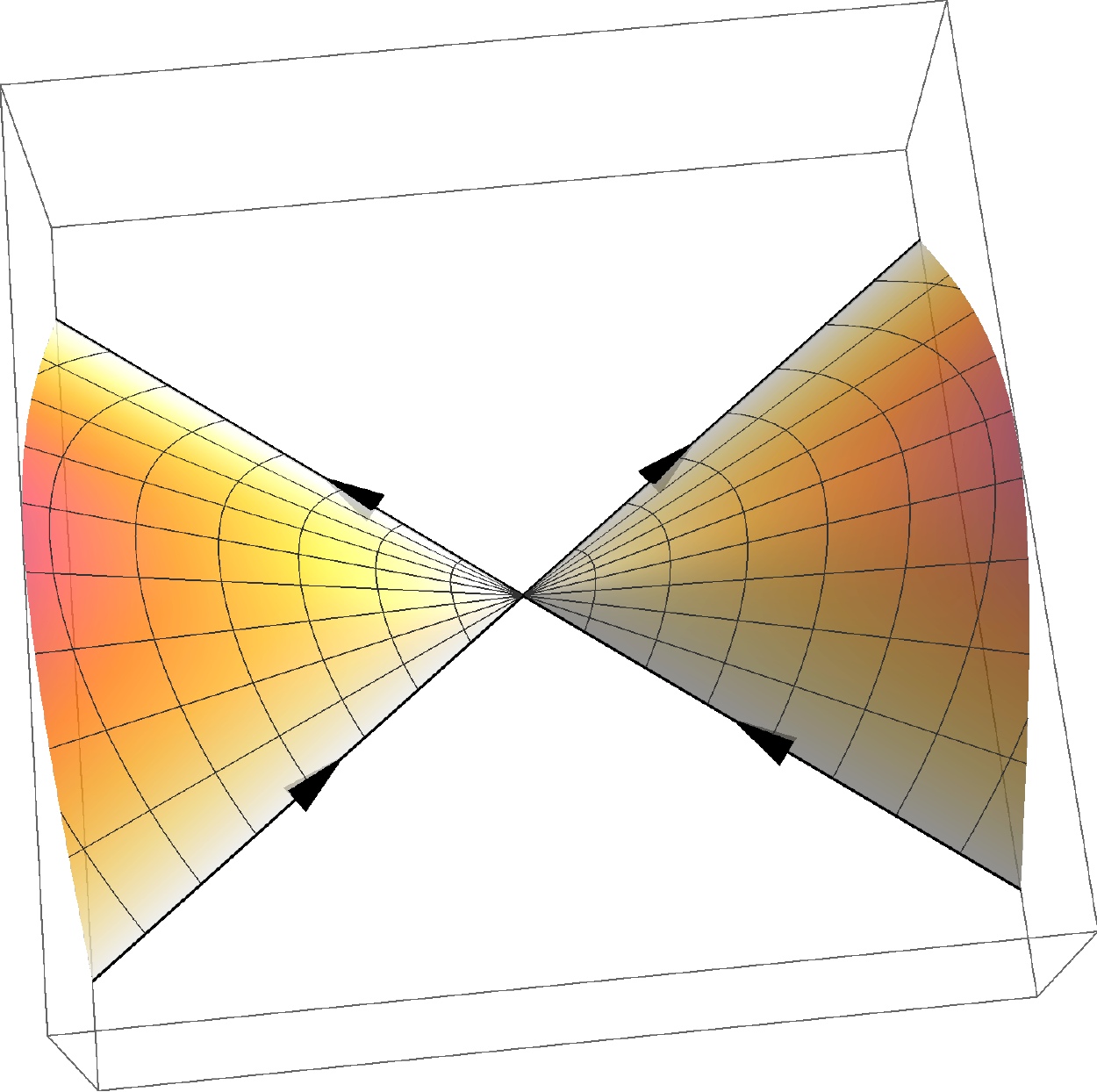} \hspace*{5mm}
\includegraphics[width=40mm]{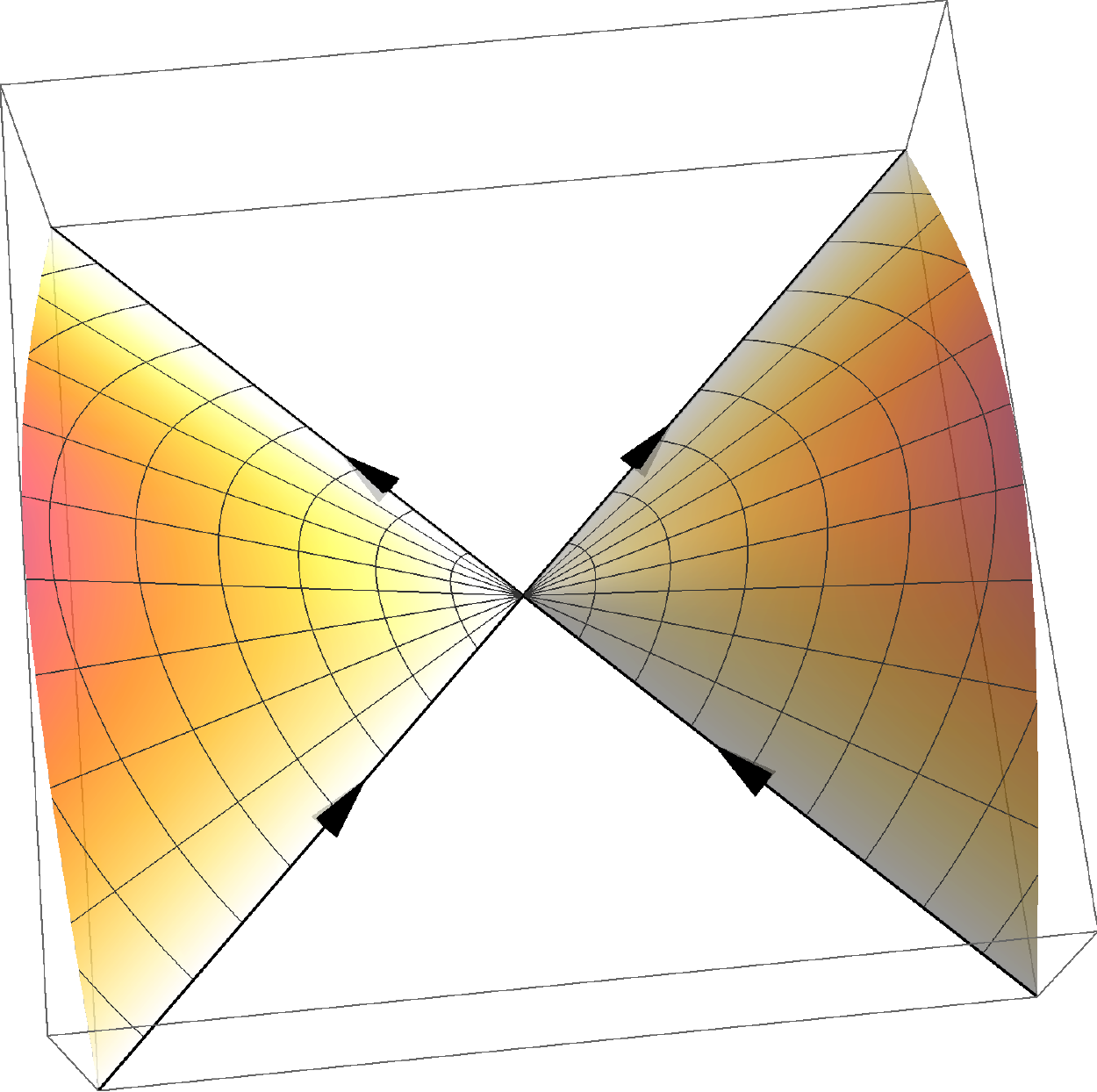} \hspace*{5mm}
\includegraphics[width=40mm]{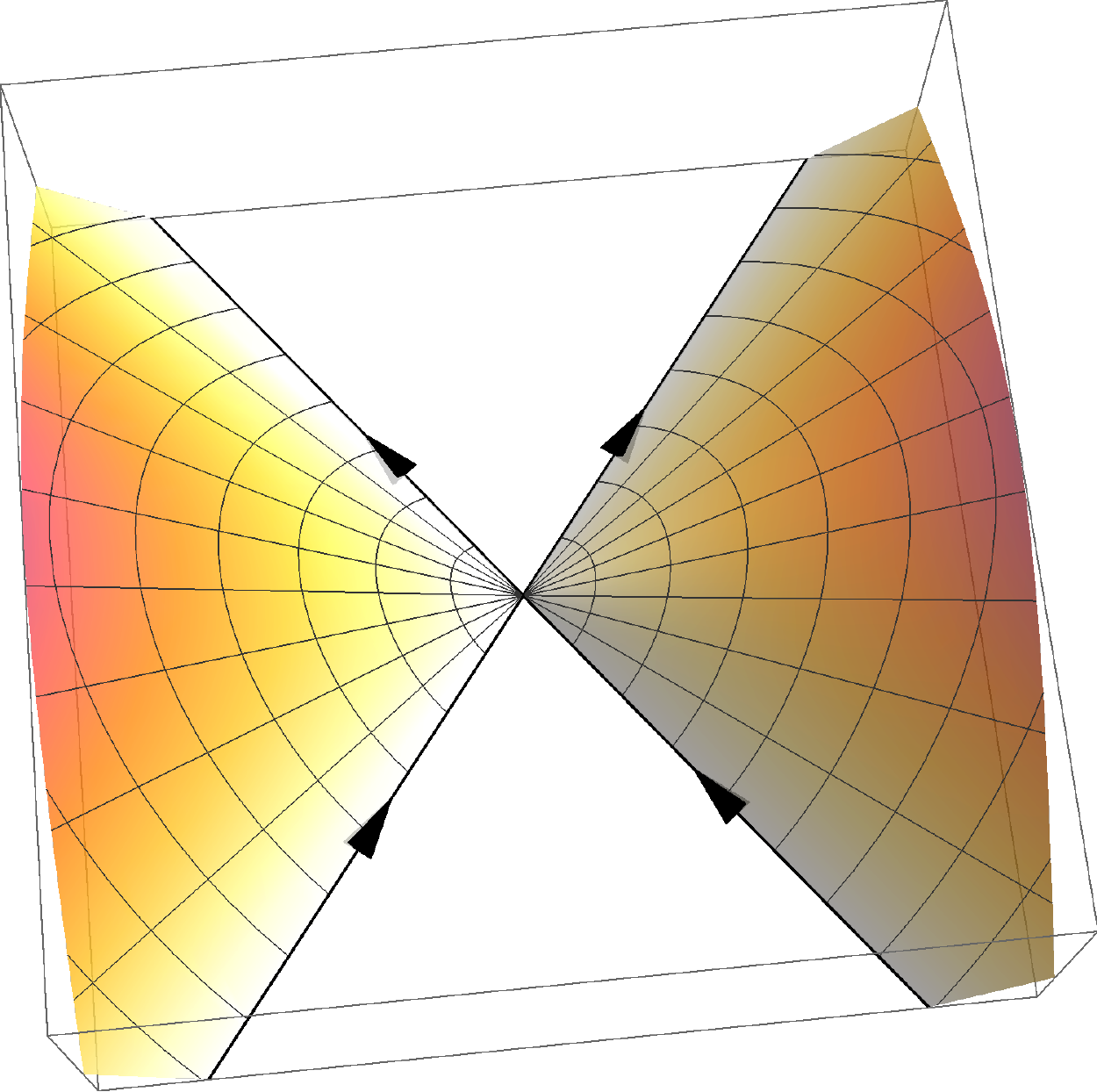} 
\caption{Sketch of the minimal surfaces for two lines intersecting at the angles $3 \pi / 8$, $\pi / 2$ and $5 \pi / 8$.}
\label{fig:minsurf1}
\end{figure}

It is instructive consider the planar limit of sending $N \to \infty$ and $g \to 0$ while keeping the 't Hooft coupling constant 
$\lambda = g^2 N$ fixed. In this limit, the leading contributions to the cross anomalous dimension $\Gacrh$ can be expressed in terms of
the cusp anomalous dimension $\Gacu$. This was noted in ref.\ \cite{Korchemskaya:1994qp} and is based on the property of large-$N$ 
factorization or vacuum dominance \cite{Makeenko:1979pb}. For two gauge invariant operators $\mathcal{O}_1$ and $\mathcal{O}_2$, we have 
\begin{align*}
\left \langle 0 \vert \mathcal{O}_1 \mathcal{O}_2 \vert 0 \right \rangle =
	\left \langle 0 \vert \mathcal{O}_1 \vert 0 \right \rangle 
	\left \langle 0 \vert \mathcal{O}_2 \vert 0 \right \rangle 
	+ \mathrm{O}\left(N^{-2}\right) , 
\end{align*}
where the operators $\mathcal{O}_i$ are assumed to be normalized in such a way that $\left \langle 0 \vert \mathcal{O}_i \vert 0 \right \rangle$ is of order $N^0$. In the present case, one can employ the large-$N$ factorization by closing the contours $C_i$ in different ways. Contracting e.g.\ $\mathcal{W}_1$ with $\vert 1 \rangle$ leads to the product the expectation values of straight Wilson lines and the contraction of $\mathcal{W}_2$ with $\vert 2 \rangle$ gives the product of two cusped Wilson loops. In the latter case, we have
\begin{align}
\mu \frac{\diff}{\diff \mu} \left \langle 
	W \left( \raisebox{-1.3mm}{\includegraphics[height=2.8ex]{Figures/curve2.pdf}} \right)
	\right \rangle ^2 =
	2 \, \Gacu (\phi , \rho , \lambda) \left \langle 
	W \left( \raisebox{-1.3mm}{\includegraphics[height=2.8ex]{Figures/curve2.pdf}} \right)
	\right \rangle ^2  .
\end{align}
In the case of straight Wilson lines, the expectation value is trivial. Contracting the renormalization group equation \eqref{rge:gacrt} 
with the color structures $\vert 1 \rangle$ and $\vert 2 \rangle$, then allows to conclude that \cite{Korchemskaya:1994qp}
\begin{align}
\Gacrh (\phi , \rho , \lambda) 
= \begin{pmatrix}
\mathrm{O}\left(N^{-2}\right) & \mathrm{O}\left(N^{-1}\right) \\
\mathrm{O}\left(N^{-1}\right) & 2 \Gacu (\phi , \rho , \lambda) 
+ \mathrm{O}\left(N^{-2}\right)
\end{pmatrix}  , 
\end{align}
and by using the relation \eqref{rel:gacr} we then have
\begin{align}
\Gacr (\phi, \rho , \lambda) 
= \begin{pmatrix}
\mathrm{O}\left(N^{-2}\right) & \mathrm{O}\left(N^0 \right) \\
\mathrm{O}\left(N^{-2}\right) & 2 \Gacu (\phi , \rho , \lambda) 
+ \mathrm{O}\left(N^{-2}\right)
\end{pmatrix}  .
\label{GammaCrossPlanar}
\end{align}
This insight allows to determine the cross anomalous dimension at strong coupling in the planar limit, where the expectation value of the Maldacena--Wilson loop is given by the area of a minimal surface in $\mathrm{AdS}_5$ ending on the respective contour on the conformal boundary \cite{Maldacena:1998im,Rey:1998ik}, 
\begin{align}
\left \langle W(C) \right \rangle =
	\exp \left( -\ft{\sqrt{\lambda}}{2\pi} \, A_{\mathrm{ren}}(C) \right) .
\end{align}
Here, $A_{\mathrm{ren}}(C)$ denotes the part of the minimal area obtained after placing a cut-off in the radial coordinate 
of $\mathrm{AdS}_5$ and subtracting the linear divergence, cf. e.g.\ ref.\ \cite{Drukker:1999zq} for details.  

In the present case, the minimal surface has to close along the cusp angle $\phi$ as shown in figure 
\ref{fig:minsurf1}, since the orientations of the lines (either both incoming or both outgoing) 
do not correspond to a minimal surface closing along the other angle. 
The two contours shown in figure \ref{fig:mixing} thus have the same associated minimal surfaces, 
such that the operators $W_1$ and $W_2$ have the same asymptotic behaviour for large $\lambda$. 
Combining this insight with the finding \eqref{GammaCrossPlanar} about the cross anomalous dimension in the planar limit, we find that
\begin{align}
\Gacr (\phi, \rho , \lambda) 
\overset{\lambda \gg 1}{=} 
\frac{\sqrt{\lambda}}{2 \pi}
\begin{pmatrix}
0 & 2 \Gacu ^{(\infty)} (\phi , \rho)  \\
0 & 2 \Gacu ^{(\infty)} (\phi , \rho) 
\end{pmatrix}  
+ \mathrm{O}\left(N^{-2}\right)
 \, .
\label{GammaCrossStrongCoupling} 
\end{align}
While the bottom-right component follows from planarity alone, the value of the top-right component of $\Gacr$ is determined via 
the AdS/CFT correspondence.
The cusp anomalous dimension at strong coupling has already been computed in ref.\ \cite{Drukker:1999zq}, consider also refs. 
\cite{Drukker:2007qr,Drukker:2011za} for more details. It can be expressed in terms of the parameters $p$ and $q$, which correspond to 
conserved quantities of the worldsheet model in appropriate coordinates. Introducing the abbreviations 
\begin{align}
b^2 &= \frac{1}{2} \left( p^2 - q^2 + \sqrt{(p^2 - q^2)^2 + 4 p^2} \right) , & 
k^2 &= \frac{b^2 \left( b^2 - p^2 \right)}{b^4 + p^2} \, , 
\end{align}
the parameters $p$ and $q$ are related to the cusp angles $\phi$ and $\rho$ by 
\begin{align}
\phi &= \pi - \frac{2 p^2}{b \sqrt{b^4 + p^2}} \left[ 
	\Pi \big( \sfrac{b^4}{b^4+p^2} \big \vert k^2 \big) - \mathbb{K} \left( k^2 \right) \right] , & 
\rho &= \frac{2 b q}{\sqrt{b^4 + p^2}} \, \mathbb{K} \left( k^2 \right)  . 	
\end{align}
Here, $\mathbb{K}$ and $\Pi$ denote complete elliptic integrals of the first and third kind, respectively, consider e.g.\ 
ref.\ \cite{Drukker:2011za} for more details. After subtracting the linear divergence, the area of the minimal surface has 
a logarithmic divergence which corresponds to the anomalous dimension at strong coupling, 
\begin{align}
A_{\mathrm{ren}} &= - \Gacu ^{(\infty)} \, \ln \left( \sfrac{L}{\varepsilon} \right) , & 
\Gacu ^{(\infty)} &= \frac{2 \sqrt{b^4 + p^2}}{b^4 + p^2} 
\left[ \frac{(b^2 + 1) p^2}{b^4 + p^2}  \mathbb{K} \left( k^2 \right) - \mathbb{E} \left( k^2 \right) \right] . 
\label{GammaCuspInfinity}
\end{align}
Here, $\varepsilon$ denotes the cut-off in $\mathrm{AdS}_5$ and $\mathbb{E}$ denotes a complete elliptic integral of the second kind. 
Since the minimal surfaces are exactly the ones occurring in the calculation of the cusp anomalous dimension, the result 
\eqref{GammaCrossStrongCoupling} should hold for the higher order coefficients of the strong-coupling expansion as well. 
This expansion is currently known up to the one-loop order for the cusp anomalous dimension \cite{Drukker:2011za}.     

\begin{figure}[t]
\centering
\includegraphics[width=100mm]{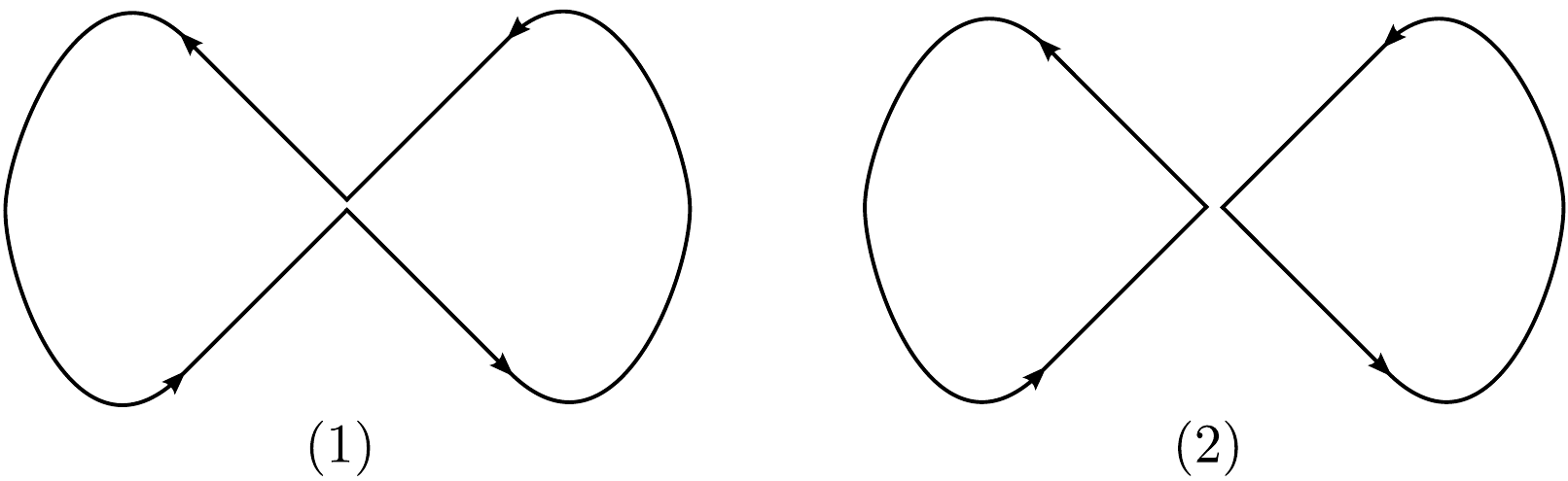}
\caption{Sketch of the contours introduced in eq.\ \eqref{cusped-lines}. The distances between the cusps merely indicate 
the ordering around the intersection point.}
\label{fig:mixing3}
\end{figure}

\begin{figure}[t]
\centering
\includegraphics[width=40mm]{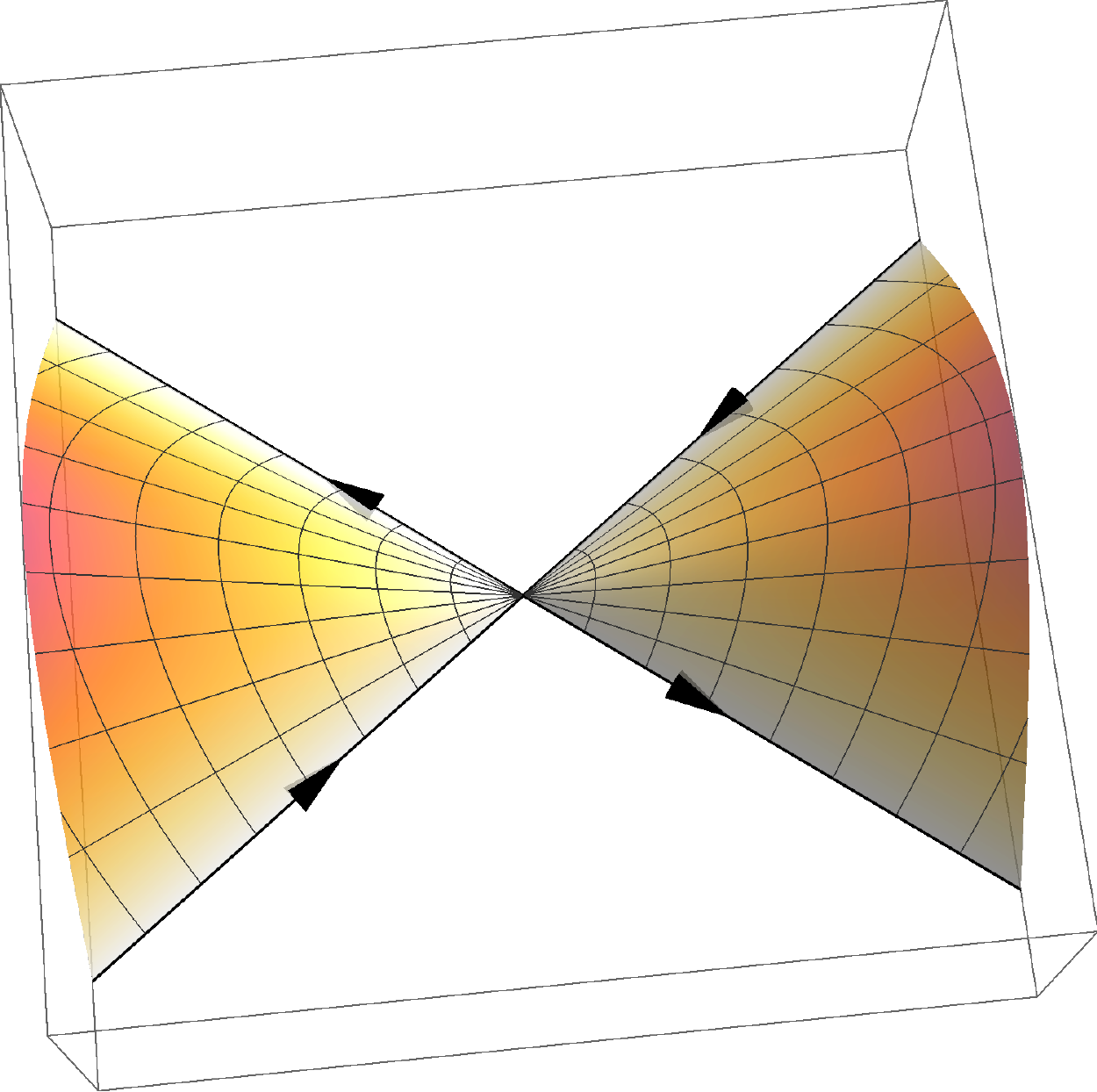} \hspace*{5mm}
\includegraphics[width=40mm]{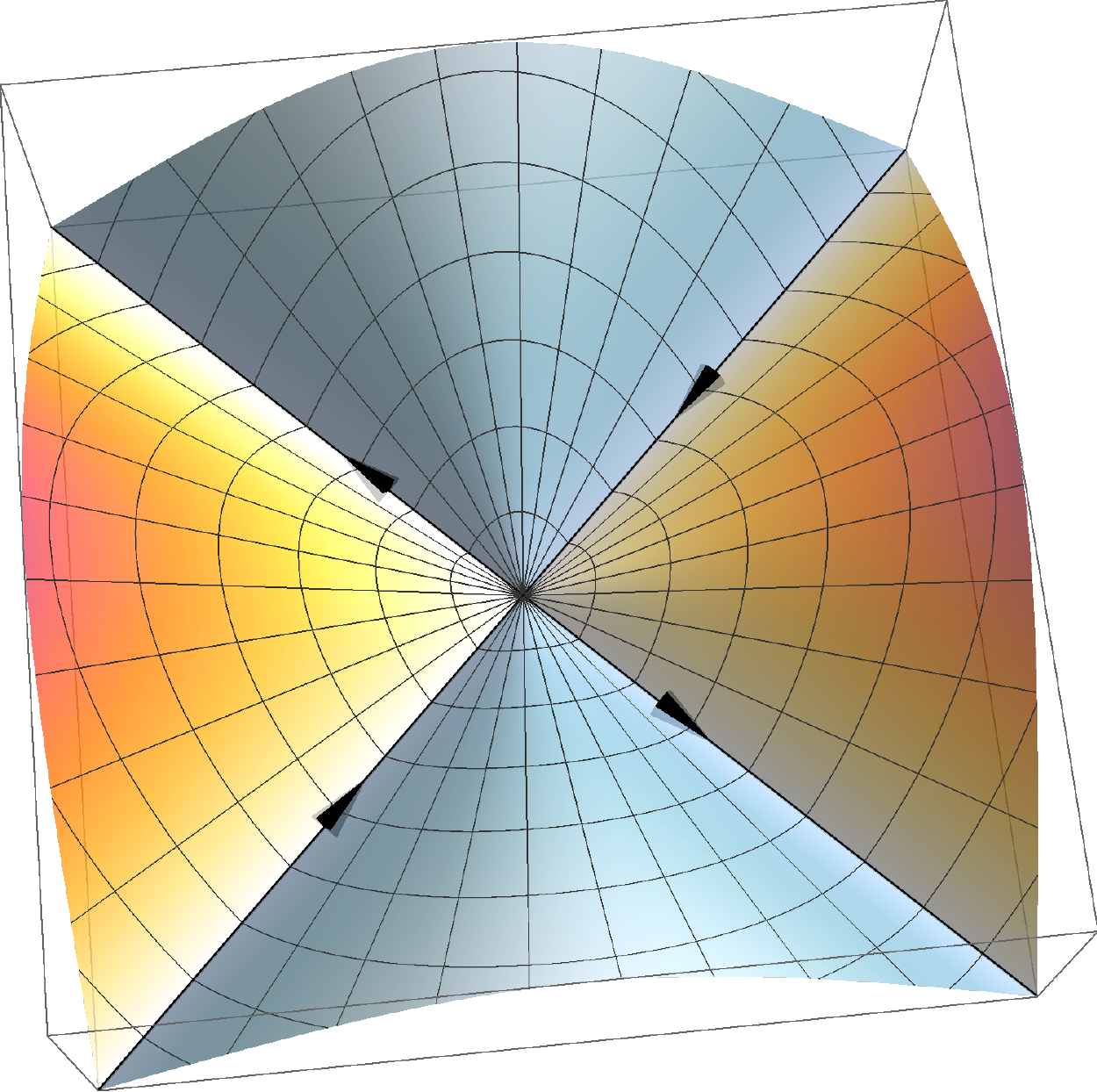} \hspace*{5mm}
\includegraphics[width=40mm]{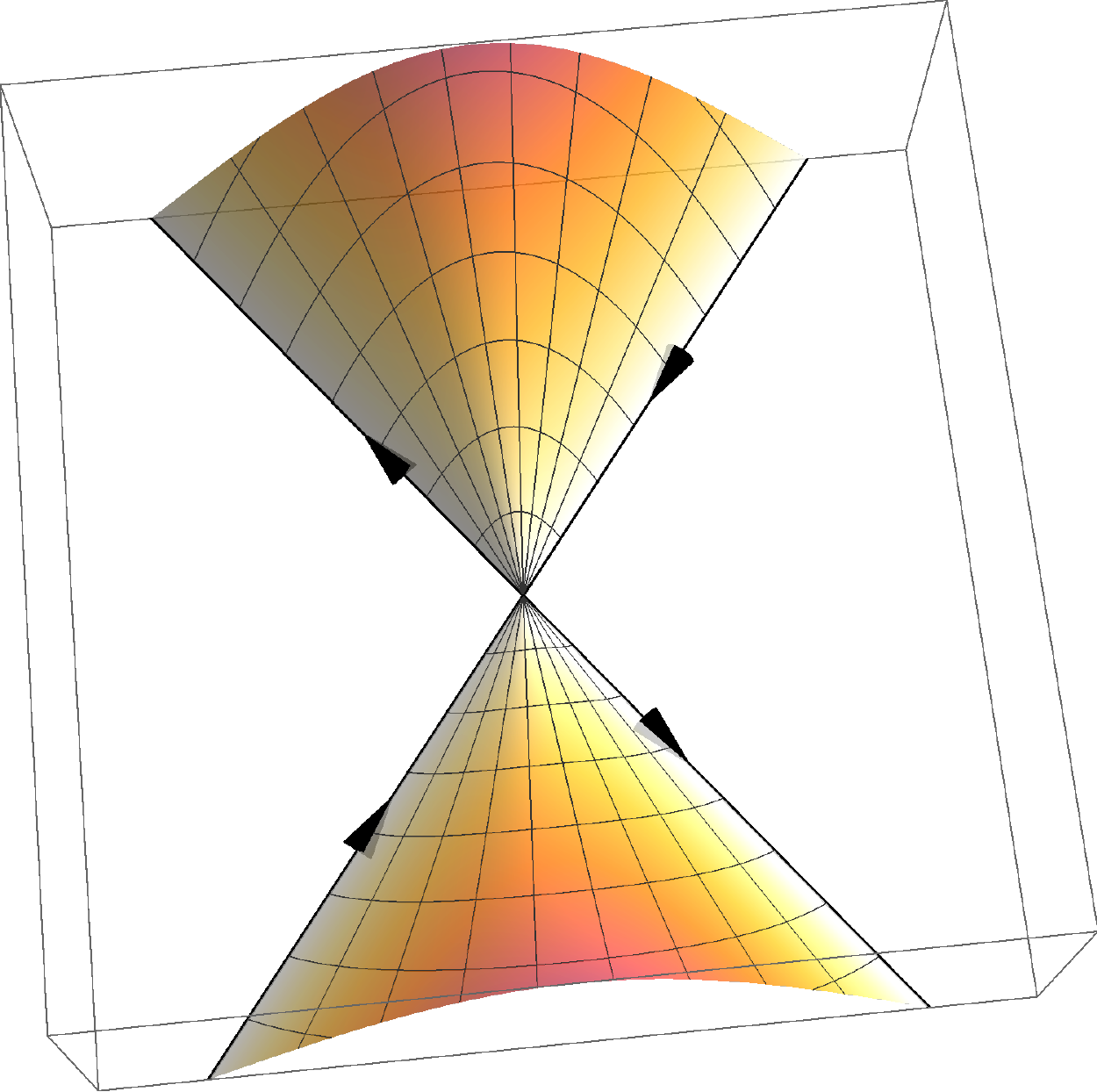}
\caption{Sketch of the minimal surfaces and phase transition for two cusped lines intersecting at the angles 
$3 \pi / 8$, $\pi / 2$ and $5 \pi / 8$.}
\label{fig:minsurf2}
\end{figure}

So far, we have restricted our attention to the case of two straight lines crossing at an angle $\phi$. 
The lines could, however, also have cusps at the intersection point. If we do not consider multiple 
self-intersections, the most general situation is given by four semi-infinite lines sharing a common starting point.  
In the following, we will restrict to the case, where the two semi-infinite lines on the right-hand side of figure 
\ref{fig:mixing} are reversed and the scalar couplings are altered in a specific way. For the first contour shown 
in figure \ref{fig:mixing3}, the two cusped lines are parametrized by
\begin{align}
( x(\sigma) , n(\sigma) )  = \begin{cases}
	(\theta(-\sigma) \, v_1 \sigma  - \theta(\sigma) \, v_2 \sigma \, , \, 
	\theta(-\sigma) \, n_1 - \theta(\sigma) \, n_2 )
	\quad &\text{for the bottom line}  \, ,\\
	(- \theta(-\sigma) \, v_2 \sigma  + \theta(\sigma) \, v_1 \sigma \, , \,
	- \theta(-\sigma) \, n_2  + \theta(\sigma) \, n_1  )   
	\quad &\text{for the top line} \, .
\end{cases}
\end{align}
In other words, we consider the sequence of lines 
\begin{align}
(v_1 , n_1) \; \; \to \; \; 
(- v_2 ,- n_2) \; \; \to \; \; 
(- v_1 ,- n_1) \; \; \to \; \; 
(v_2 , n_2) \, , 
\label{cusped-lines} 
\end{align}
We denote the respective Wilson loop correlators by $\widetilde{W}_i$ and 
the associated cross anomalous dimension by $\Gacrt$. 
The set-up contains additional cusps with angles $\pi - \phi$ and $\pi - \rho$ at the intersection 
point, altering the cross anomalous dimension. Repeating the analysis of ref.\ \cite{Korchemskaya:1994qp} for the planar limit, we 
find that the contraction of the correlator $\widetilde{\mathcal{W}}_1$ with the color structure $\one$ gives the product of two cusped Wilson lines, 
such that the cross anomalous dimension $\Gacrt$ takes the form 
\begin{align}
\Gacrt (\phi, \rho , \lambda) 
= \begin{pmatrix}
2 \Gacu (\pi - \phi ,\pi - \rho , \lambda) + \mathrm{O}\left(N^{-2}\right) & \mathrm{O}\left(N^0 \right) \\
\mathrm{O}\left(N^{-2}\right) & 2 \Gacu (\phi , \rho , \lambda) 
+ \mathrm{O}\left(N^{-2}\right)
\end{pmatrix} \, .
\label{GammaCrossTildePlanar}
\end{align}
For the minimal surfaces in this case, we have two competing saddle points, since the surface can close along either cusp angle, 
see figure \ref{fig:minsurf2}. The logarithmic divergence of the minimal surfaces is then given by 
$\Gacu ^{(\infty)}  (\phi, \rho)$ or $\Gacu ^{(\infty)} (\pi - \phi, \pi - \rho)$, respectively, the smaller area corresponding 
to the larger anomalous dimension. The correlator $\widetilde{W}_1$ thus undergoes a Gross--Ooguri phase transition at the point where 
the two anomalous dimensions match. A transition of this kind has been studied in the early days of the AdS/CFT correspondence 
for the correlation function of two circular 
Maldacena--Wilson loops \cite{Gross:1998gk,Zarembo:1999bu,Olesen:2000ji,Zarembo:2001jp} and more recently for antiparallel lines 
in the presence of a defect \cite{Preti:2017fhw}. 
The angle $\phi$ at which the phase transition occurs depends on the value of $\rho$ and can at least numerically 
be determined from the infinite-coupling limit of the cusp anomalous dimension \eqref{GammaCuspInfinity}. For $\rho = \pi /2$, it must 
occur at $\phi = \pi /2$ as depicted in figure \ref{fig:minsurf2}.     

Since the top-left component of the cross anomalous dimension is fixed by planarity, we must adapt 
the top-right component to the divergence of the area. This gives the following result for the cross anomalous dimension 
$\Gacrt$ in the limit of infinite coupling: 
\begin{align}
\Gacrt (\phi, \rho , \lambda) \overset{\lambda \gg 1}{=}
	\frac{\sqrt{\lambda}}{\pi} \begin{pmatrix}
	\Gacu ^{(\infty)} (\pi - \phi, \pi - \rho)  & 
	\text{max} \big \lbrace 0 ,  \Gacu ^{(\infty)}  (\phi, \rho) 
		- \Gacu ^{(\infty)}  (\pi - \phi, \pi - \rho ) \big \rbrace \\
	0 & \Gamma _{\text{cusp}} ^{(\infty)} (\phi) 
\end{pmatrix} .
\end{align}
The Gross--Ooguri transition of the correlator $\widetilde{W}_1$ is hence mirrored in the top-right component of 
the cross anomalous dimension matrix. The transition occurs for strictly infinite coupling; for large values of the 
coupling constant, the dependence of the string partition function on the cusp angles is smoothened over to an expansion
around both saddle points, 
\begin{align*}
\widetilde{W}_1 \simeq c_1 \exp{ \left( \ft{\sqrt{\lambda}}{2 \pi} \Gacu ^{(\infty)}  (\phi, \rho) \ln (\mu L) \right) } 
	+ c_2 \exp{ \left( \ft{\sqrt{\lambda}}{2 \pi} \Gacu ^{(\infty)}  (\pi - \phi, \pi -  \rho) \ln (\mu L) \right) } . 
\end{align*}
Here, $\mu$ denotes the generalized renormalization scale. 

\section{Weak Coupling -- One Loop}
\label{sec:OneLoop}
\begin{figure}[t]
\centering
\includegraphics[width=107mm]{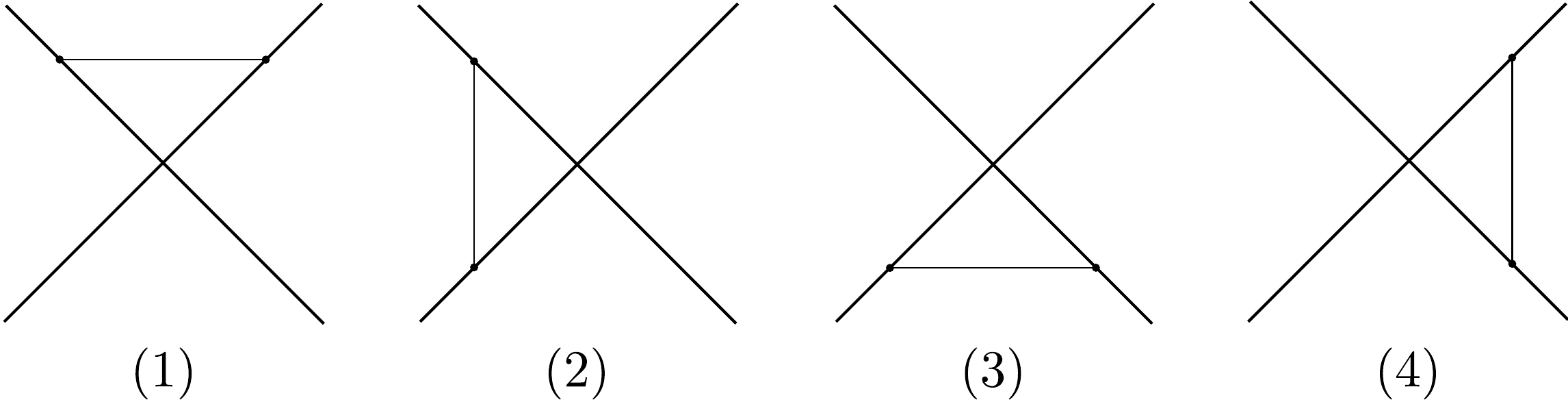}
\caption{Diagrams at the one-loop level. The straight line denotes a combination of gluon and scalar propagators.}
\label{fig:one-loop}
\end{figure}
We now turn to the weak-coupling calculation of the cross anomalous dimension, beginning with the one-loop level. Following ref.\ \cite{Erickson:2000af}, we employ dimensional reduction to regularize divergences. Dimensional reduction is a version of dimensional regularization in which $\mathcal{N} \! =  4$ supersymmetric Yang--Mills theory in $\mathrm{D}=4 - 2 \epsilon$ dimensions is viewed as the theory obtained from the dimensional reduction of ten-dimensional $\mathcal{N}=1$ supersymmetric Yang--Mills theory to D dimensions. The regularized theory hence has a D-component vector field $A_\mu ^a$ as well as $10 - \mathrm{D}$ scalar fields $\Phi _I ^a$, see \cite{Siegel:1979wq} for more details. For the renormalization of the UV divergences we choose the minimal subtraction scheme.   

At the one-loop level, the regularization only appears through the alteration of the two-point functions. The propagator in momentum space is unaltered but the Fourier transformation is carried out in D dimensions. For the scalar propagator, we then have \cite{Erickson:2000af}
\begin{align}
D(x) = g^2 \mu^{2 \epsilon} \, \frac{\Gamma (1- \epsilon )}{4 \pi^{2-\epsilon} } \, 
	\frac{1}{ \left( x^2 \right) ^{1- \epsilon} } \, ,
\end{align}
and the two-point functions
\begin{align}
\left \langle A_\mu ^a (x_1) \, A_\nu ^b (x_2) \right \rangle = 
	\delta _{\mu \nu} \, \delta^{a b} D(x_1 - x_2) 
	\, , \qquad 
	\left \langle \Phi_I ^a (x_1) \, \Phi_J ^b (x_2) \right \rangle = 
	\delta _{I J} \, \delta^{a b} D(x_1 - x_2) \, .
\end{align}
In dealing with the Maldacena--Wilson loop it is natural to introduce the abbreviations
\begin{align}
A ( \tau_1 ) &= A _\mu (x_1) \, \dot{x}_1 ^\mu \, , &
	\Phi (\tau_1) &= \Phi_I (x_1) \, n_1 ^I \, \lvert \dot{x}_1 \rvert \, .
\end{align}
The relevant two-point functions for the Maldacena--Wilson loop are then given by 
\begin{align*}
\left \langle 
	\Big( A ^a ( \tau_1 ) +  i \Phi ^a (\tau_1)  \Big)
	\Big( A ^b ( \tau_2 )  +  i \Phi ^b (\tau_2)  \Big)
	\right \rangle
	= g^2 \mu ^{2 \epsilon} \, \frac{\Gamma (1- \epsilon )}{4 \pi^{2-\epsilon} } \, 
	\frac{\dot{x}_1 \dot{x}_2 - n_1 n_2 \lvert \dot{x}_1 \rvert \lvert \dot{x}_2 \rvert }
	{\left[ (x_1 - x_2)^2 \right] ^{1- \epsilon}} \, \delta^{a b} \, .
\end{align*}
Since the $\mathrm{S}^5$-vector $n$ is constant along the lines, the above correlator vanishes if $\tau_1$ and $\tau_2$ are on the same line. We thus only need to consider the four diagrams depicted in figure \ref{fig:one-loop}. Following ref.\ \cite{Korchemskaya:1994qp}, we separate the contribution of each diagram into a color factor and a kinematic factor in such a way that
\begin{align}
\mathcal{W}_a 
	&= \sum \limits _n \lambda ^n \, \mathcal{W}_a  ^{\,(n)}  , &
	\mathcal{W}_a  ^{\,(1)} &= \sum \limits _m \ft{1}{N}  F^1 _m \, C^1 _{a, m} \, .
\end{align}
Here, $F^1 _m$ denotes the kinematic factor associated to diagram $m$ of figure \ref{fig:one-loop}, whereas $C^1 _{a, m}$ denotes the color factor associated to this diagram given the path-ordering of $\mathcal{W}_a$ around the intersection point. The convention to include a factor of $N^{-1}$ is a consequence of expanding in the 't Hooft coupling constant $\lambda = g^2 N$. For diagram 1, we find the color factors
\begin{align*}
C^1_{1,1} &= T^a_{j j^\prime} \, T^a_{i i^\prime} 
	= \ft{1}{2} \left( \delta_{i j^\prime} \, \delta_{j i^\prime}
	- \ft{1}{N} \delta_{i i^\prime} \, \delta_{j j^\prime} \right)
	= \ft{1}{2} \two - \ft{1}{2N} \one \, , \\ 
C^1_{2,1} &= T^a_{j i^\prime} \, T^a_{i j^\prime} 
	= \ft{1}{2} \left( \delta_{i i^\prime} \, \delta_{j j^\prime}
	- \ft{1}{N} \delta_{j i^\prime} \, \delta_{i j^\prime} \right)
	= \ft{1}{2} \one - \ft{1}{2N} \two \, .
\end{align*}
The color factors for the other diagrams are collected in appendix \ref{app:Color}. For the kinematic factor associated to diagram 1 we find
\begin{align*}
F^1 _1 &= - \frac{\Gamma (1 - \epsilon)}{4 \pi ^{2 - \epsilon} } \, g^2 \mu ^{2 \epsilon} 
	\int \limits _0 ^L \diff \tau_1 \, \diff \tau_2 \,
	\frac{\cos \phi - \cos \rho}
	{\left[ \tau_1 ^2 + \tau_2 ^2 - 2 \tau_1 \tau_2 \cos \phi \right]^{1- \epsilon} } 
	= \frac{g^2 \mu^{ 2 \epsilon}}{\epsilon} \, 
	\left(\cos \phi - \cos \rho \right) 	
	I(\pi - \phi )
	\, .
\end{align*}
Here, we have introduced the angles $\cos \phi = v_1 v_2$ and $\cos \rho = n_1 n_2$. The kinematic factors for the other diagrams can be related to $F^1_1$. For $F^1_2$, we find 
\begin{align*}
F^1 _2 &= - \frac{\Gamma (1 - \epsilon)}{4 \pi ^{2 - \epsilon} } \,g^2  \mu ^{2 \epsilon} 
	\int \limits _{-L} ^0  \diff \tau_1 \, 
	\int \limits _0 ^L  \diff \tau_2 \,
	\frac{\cos \phi - \cos \rho}
	{\left[ \tau_1 ^2 + \tau_2 ^2 - 2 \tau_1 \tau_2 \cos \phi \right]^{1- \epsilon} } \\
	&= - \frac{\Gamma (1 - \epsilon)}{4 \pi ^{2 - \epsilon} } \, g^2 \mu ^{2 \epsilon} 
	\int \limits _0 ^L \diff \tau_1 \, \diff \tau_2 \,
	\frac{\cos \phi - \cos \rho}
	{\left[ \tau_1 ^2 + \tau_2 ^2 + 2 \tau_1 \tau_2 \cos \phi \right]^{1- \epsilon} } 
	= \frac{g^2 \mu^{ 2 \epsilon}}{\epsilon} \, 
	\left(\cos \phi - \cos \rho \right) 	
	I (\phi ) 
	\, .
\end{align*}
The relation found above is a consequence of the behaviour of the Maldacena--Wilson loop under a reflection $v_2 \to -v_2$ or $\phi \to (\pi - \phi)$ in terms of the angle between $v_1$ and $v_2$. The reflection maps the curve to itself but changes its direction. Under such a reflection, the Wilson line integrals corresponding to the same part of the curve are related by
\begin{align*}
\int \limits _0 ^L A_\mu(x) \dot{x} ^\mu \diff \tau &\to 
	  - \int \limits _{-L} ^0 A_\mu(x) \dot{x} ^\mu \diff \tau \, , &
\int \limits _0 ^L \Phi_I(x) \lvert \dot{x} \rvert n^I \diff \tau &\to 
	  \int \limits _{-L} ^0 \Phi_I(x) \lvert \dot{x} \rvert n^I \diff \tau \, .	  
\end{align*}  
Due to the different sign between the two contributions, we relate $F^1 _2( \phi , \rho )$ to 
$F^1 _1(\pi - \phi , \pi - \rho)$, i.e.\ with an altered scalar coupling. 
Equivalently, we may factor out the scalar product factor $( \cos \phi - \cos \rho)$, which accounts for the different signs. We can relate the kinematic factors for the two additional diagrams to the ones discussed above by reflecting both $v_1$ and $v_2$. Then we see that 
\begin{align*}
F^1 _3 &= F^1 _1 \, , &
F^1 _4 &= F^1 _2 \, .
\end{align*} 
The cross anomalous dimension at the one-loop level is thus encoded in the function $I_0 (\phi)$, which also describes the cusp anomalous dimension. 

We then turn to the evaluation of $I(\phi) = I_0 (\phi) + \epsilon I_1 (\phi) + \LO ( \epsilon^2) $. For the anomalous dimension, we need only compute the pole coefficient $I_0 (\phi)$. The divergence stems from the limit of both $\tau_1$ and $\tau_2$ approaching the cusp and can be isolated by substituting
\begin{align*}
\tau_1 &= \kappa z \, , &
\tau_2 &= \kappa \zb \, , 
\end{align*}
where we have abbreviated $\zb = 1 -z $. For the coordinate ranges we note that $z \in [0,1]$ and $\kappa \in [0, f(z)]$. The divergence is now captured in the limit $\kappa \to 0$ such that the precise form of $f(z)$ is not relevant for the pole coefficient $I_0 (\phi)$ and thus we take $\kappa \in [0, 1]$ as well. Then we get
\begin{align}
I_0 (\phi) &= 
	- \frac{\epsilon }{4 \pi^2} 
	\int \limits _0 ^1 \diff z \,
	\frac{1}{\left( z^2 + \zb^2 + 2 z \zb \cos \phi \right)} 
	\int \limits _0 ^1 \diff \kappa \, \kappa ^{-1+2 \epsilon} 
	= - \frac{1}{8 \pi^2} 
	\int \limits _0 ^1 \diff z \,
	\frac{1}{\left( z^2 + \zb^2 + 2 z \zb \cos \phi \right)} \nn \\
	&= - \frac{\phi}{8 \pi^2 \sin \phi} \, .
\label{I0-result}	
\end{align}
Here, we have obtained the remaining integral by decomposing into partial fractions. 

With all kinematic factors under control, we can then determine the cross anomalous dimension matrix. Using the color factors given in appendix \ref{app:Color}, we find 
\begin{align}
\begin{pmatrix}
	\mathcal{W}_1 ^{\, (1)}  \\
	\mathcal{W}_2 ^{\, (1)} 
\end{pmatrix}  =
	\frac{\mu^{2 \epsilon} ( \cos \phi - \cos \rho )}{N^2 \, \epsilon}
	\begin{pmatrix}
		- (I(\phi) + I(\pi - \phi)) & N (I(\phi) + I(\pi - \phi)) \\
		N \, I(\pi - \phi) & N^2 \, I(\phi) - ( I(\phi) + I(\pi - \phi) )	
	\end{pmatrix}
	\begin{pmatrix}
	\one \\
	\two
	\end{pmatrix} .
\end{align}
The $Z$-factor is correspondingly given by
\begin{align}
\widehat{Z} _{\mathrm{cross}} ^{(1)} (\phi, \rho) 
	= \frac{( \cos \phi - \cos \rho )}{N^2 \epsilon} 
	\begin{pmatrix}
		I_0(\phi) + I_0(\pi - \phi) & - N (I_0(\phi) + I_0(\pi - \phi)) \\
		-N I_0(\pi - \phi) & - N^2 \, I_0(\phi) + ( I_0(\phi) + I_0(\pi - \phi) ) 	
	\end{pmatrix} .
\end{align}
The anomalous dimension can then be determined from the renormalization group equation \eqref{rge:gacrt}. Additionally, we make use of the relation \eqref{rel:gacr} to determine the anomalous dimension for $W_a$ from the one for $\mathcal{W}_a$. Then we get  
\begin{align}
\Gacr ^{(1)} (\phi, \rho ) = 
	\frac{2 ( \cos \phi - \cos \rho )}{N^2}
	\begin{pmatrix}
		- I_0(\phi) - I_0(\pi - \phi) & N^2 (I_0(\phi) + I_0(\pi - \phi) ) \\
		I_0(\pi - \phi) & N^2 \, I_0(\phi) - ( I_0(\phi) + I_0(\pi - \phi) )	
	\end{pmatrix} .
\end{align}
It is clear that the function $I_0(\phi)$ also describes the cusp anomalous dimension, for which one only has the kinematic factor $F^1 _2$. Combining this factor with the color factor $\tr ( T^a T^a) = \half (N^2 - 1)$ gives the cusp anomalous dimension at the one-loop level, 
\begin{align}
\Gacu ^{(1)} (\phi, \rho ) = \frac{N^2-1}{N^2}
	\left( \cos \phi - \cos \rho \right) I_0(\phi) 
	=  \frac{N^2-1}{N^2} \, \gacu ^{(1)} (\phi, \rho )  \, .
\label{I0-Gamma}
\end{align}
The contribution $I_0(\pi - \phi)$ from the color factor $F^1 _2$ can 
also be expressed in terms of the cusp anomalous dimension,   
\begin{align*}
\left( \cos \phi - \cos \rho \right) \, I_0(\pi - \phi) = 
	- \gacu ^{(1)} (\pi - \phi, \pi - \rho ) = - \gacur ^{(1)} (\phi, \rho ) \, .
\end{align*}
We can thus --- as expected at the one-loop order --- express the cross anomalous dimension in terms of 
the cusp anomalous dimension, 
\begin{align}
\Gacr ^{(1)} =  2 \begin{pmatrix}
	0 & \gacu^{(1)} - \gacur^{(1)} \\
	0 & \gacu^{(1)}
	\end{pmatrix}
	- \frac{2}{N^2} 
	\begin{pmatrix}
	\gacu^{(1)} - \gacur^{(1)} & 0 \\ 
	\gacur^{(1)} & \gacu^{(1)} - \gacur^{(1)}
	\end{pmatrix} . 
\end{align}

Let us now consider the configuration shown in figure \ref{fig:mixing3}. The scalar couplings for this contour are chosen 
in such a way that diagrams with propagators on opposite edges do not contribute due to cancellations between gluon and 
scalar propagators. We thus again consider the diagrams shown 
in figure \ref{fig:one-loop} and note that the kinematic factors $\Ft ^1 _i$ can be related to the kinematic factors 
$F ^1 _i$ in a simple way. Consider e.g.\ the kinematic factor associated to the first diagram,
\begin{align}
\Ft _1 ^1 = - \frac{\Gamma ( 1 - \epsilon) }{4 \pi ^{2 - \epsilon}} g^2 \mu^{2 \epsilon}
	\int \limits _0 ^L \diff \tau_1 \, \diff \tau_2 \,
	\frac{- \cos \phi + \cos \rho}
	{\left[ \tau_1 ^2 + \tau_2 ^2 - 2 \tau_1 \tau_2 \cos \phi \right]^{1- \epsilon} } = -F_1 ^1 \, . 
\end{align}
In a similar fashion, we find that the kinematic factor associated to the second diagram is the same, 
\begin{align}
\Ft _2 ^1 = F _2 ^1 \, . 
\end{align}
It thus only remains to calculate the color factors, which is a straight-forward exercise, 
\begin{align}
\Ct ^1 _{1,1} &= \frac{N^2-1}{2N} \one \, , & 
\Ct ^1 _{1,2} &= \half \two - \ft{1}{2N} \one \, ,  \\ 
\Ct ^1 _{2,1} &= \half \one - \ft{1}{2N} \two \, , & 
\Ct ^1 _{2,2} &= \frac{N^2-1}{2N} \two \, . 
\end{align}
Summing up these contributions, we find the cross anomalous dimension for the correlators $\widetilde{W}_1$ and 
$\widetilde{W}_2$ to be given by
\begin{align}
\Gacrt ^{(1)} = 2 \begin{pmatrix}
	\gacur ^{(1)}  & 
	\gacu ^{(1)}  \\
	0 & \gacu ^{(1)} 
	\end{pmatrix}
	- \frac{2}{N^2} 
	\begin{pmatrix}
	\gacu ^{(1)} + \gacur ^{(1)} & 0 \\ 
	- \gacur ^{(1)} & \gacu ^{(1)} + \gacur ^{(1)}	
	\end{pmatrix} . 
\end{align}

\section{Weak Coupling -- Two Loops}
\label{sec:TwoLoop}

Before we turn to the two-loop calculation of the cross anomalous dimension, let us comment on some organizing principles 
that can be applied. In the calculation of both the cross and the cusp anomalous dimension, one acquires a factor 
$\xi = (\cos \phi - \cos \rho)$ for each connection of one Wilson line to another. It is hence convenient to expand the 
anomalous dimensions in powers of $\xi$ and discuss the expansion coefficients separately, i.e.\ to consider the expansion 
\begin{align}
\Gacu ^{(n)} (\phi, \rho ) = 
	\sum \limits _{k=1} ^n 
	\xi ^k \, \Gacu ^{(n, k )} (\phi) \, .
\label{GammaCuspExpansion}
\end{align} 
Since the $\xi$-factors contain the entire dependence of the anomalous dimension on $n_1 n_2 = \cos \rho$, cancellations 
between the different coefficients do not occur. The leading coefficient at each order contains the contributions of all 
ladder-like diagrams.

Let us furthermore comment on the structure of the $Z$-factor and anomalous dimension at higher loop orders in general. 
The correlators of Wilson lines are known to exponentiate,
\begin{align}
\begin{pmatrix}
	\mathcal{W}_1   \\
	\mathcal{W}_2  
\end{pmatrix}  = 
 	\exp \left( \sum \limits _{n=1} ^\infty  \lambda ^n \, w^{(n)} ( \epsilon , \mu , N )  \right)  
 	\begin{pmatrix}
		\one \\
		\two
	\end{pmatrix} .
\label{Wexp}	
\end{align}
In the case of the cusp anomalous dimension, a similar exponentiation formula had already been employed in the two-loop calculation of ref.\ \cite{Korchemsky:1987wg}. There, the exponent is given by summing over a subset of diagrams with modified color factors. The respective diagrams are called color-connected or maximally non-Abelian. In the case of the cross or soft anomalous dimension matrix, the exponent has only recently been constructed in ref.\ \cite{Gardi:2010rn,Mitov:2010rp}, see also ref.\ \cite{White:2015wha} for a pedagogical introduction. Here, the exponent involves not just the sum over a subset of diagrams but contains combinations of diagrams known as webs, in which the different diagrams are weighted by the coefficients of the web-mixing matrix. 

One aspect that can be observed from the exponentiation formula \eqref{Wexp} is that due to the matrix structure of the coefficients $w^{(n)}$, the $Z$-factor is not simply given by exponentiating the poles of the coefficients $w^{(n)}$, but involves commutator terms which are dictated by the Baker--Campbell--Hausdorff formula. This is also the case for the anomalous dimension matrix, which contains the commutator $[w^{(1,-1)} , w^{(1,0)}]$ at the two-loop order. Here, $w^{(1,-1)}$ and $w^{(1,0)}$ denote the pole and finite coefficient of $w^{(1)}$ in the $\epsilon$-expansion, respectively. 

\begin{figure}
\centering
\includegraphics[width=160mm]{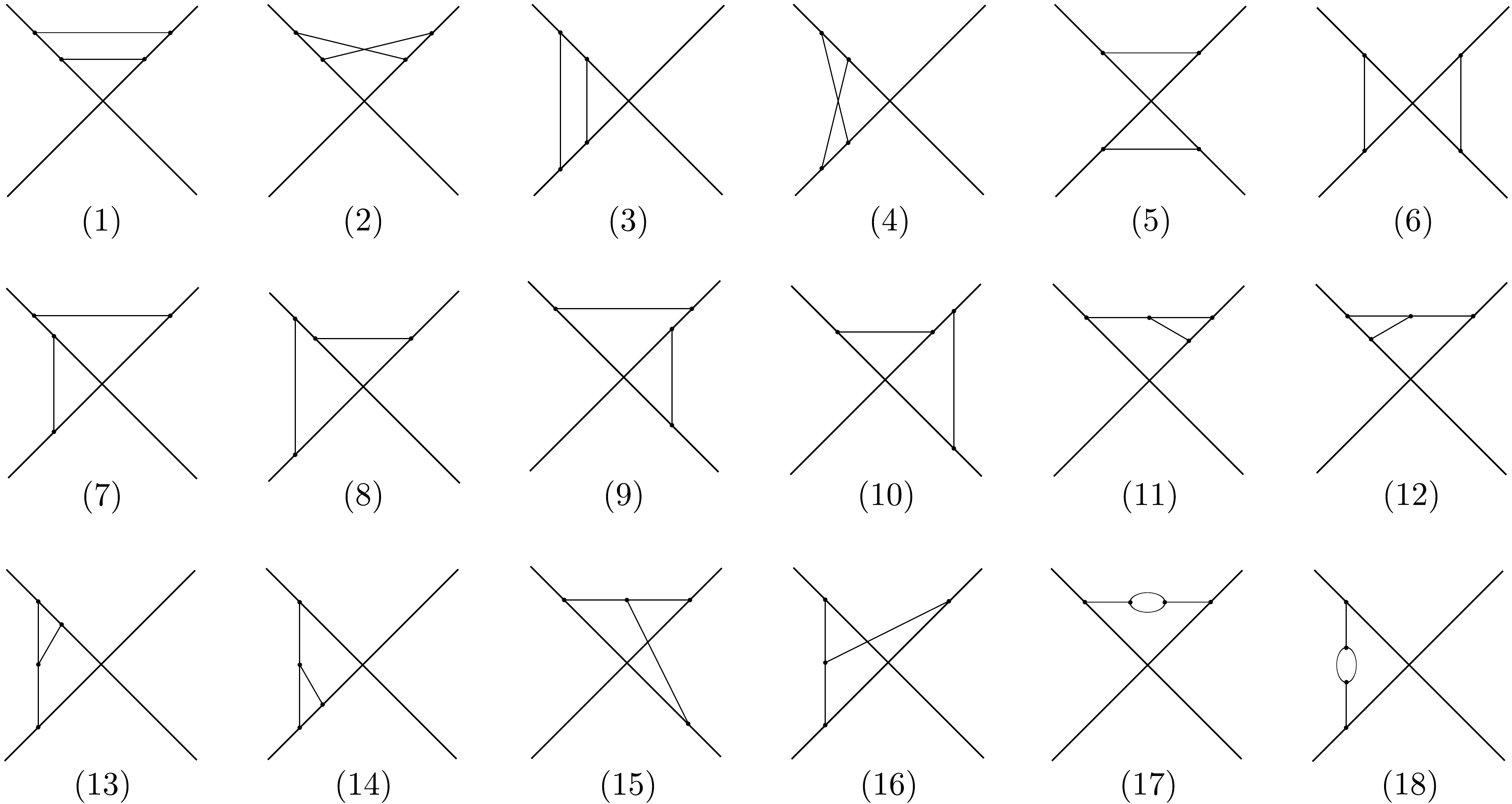}
\caption{All diagrams at the two-loop level.}
\label{fig:two-loop}
\end{figure}

While the exponentiation described above provides interesting insights into the structure of the cross anomalous dimension at two or more loops, it does not facilitate the two-loop calculation described in this paper considerably. In fact, we carry out the two-loop calculation below without referring to the exponentiation formula \eqref{Wexp}, similar to the QCD calculation carried out in ref.\ \cite{Korchemskaya:1994qp}. The diagrams for the two-loop calculation are collected in figure \ref{fig:two-loop}. Here, we have already accounted for the relations established by applying a reflection to both lines and left out diagrams which are related to given diagrams in this way. This implies that all kinematic factors except for $F^2 _5$ and $F^2 _6$ have to be accompanied by a combinatorial factor of 2.

The exponentiation of the pole terms can be seen explicitly on the level of diagrams by relating combinations of kinematic factors at the two-loop level to products of the kinematic factors encountered at the one-loop level. A guiding principle in this approach is that any occurring double pole should be contained in a product of one-loop diagrams, such that the remaining combinations of kinematic factors only contain single poles. We should hence find relations for all diagrams involving a subdivergence. Apart from the ladder-like diagrams containing two gluon or scalar exchanges between the Wilson lines, we also discuss relations between the three-vertex and self-energy diagrams. Cancellations between these contributions have been observed in the computation of the 
$\mathcal{N} \!=4$ SYM cusp anomalous dimension in ref.\ \cite{Makeenko:2006ds} and are also present here. With the relations between the kinematic factors established, we will find that the two-loop calculation of the cross anomalous dimension can be reduced to the computation of three independent combinations of kinematic factors.    

As for the one-loop calculation, it is clear that some --- but no longer all --- of the kinematic factors occur also in the calculation of the cusp anomalous dimension at the two loop level. We make this relation explicit and identify the respective kinematic factors with parts of the cusp anomalous dimension. Then the calculation of the cross anomalous dimension requires only the calculation of the commutator term as well as one additional combination of kinematic factors known as the two-gluon 
exchange web.
\subsection{Relations Between Diagrams}
Many of the diagrams depicted in figure \ref{fig:two-loop} can be related to each other or to products of one-loop diagrams. We work out these relations below, starting with the ladder-like diagrams.
\vspace*{-2mm} 
\paragraph{Ladder-like Diagrams.} $\mbox{}$\\
We begin by considering the diagrams $F^2 _1$ and $F^2 _2$. Employing the abbreviations 
$\theta_{ij} = \theta(\tau_i - \tau_j)$ and $D_{ij} = D(\tau_i v_1 - \tau_j v_2)$, we have
\begin{align*}
F^2 _1  &= \xi^2 \! \int \limits _0 ^L \diff^4 \tau \, 
	\theta_{31} \, \theta_{42} \, 
	D_{12} \, 
	D_{34} \, , &
F^2 _2  &= \xi^2 \! \int \limits _0 ^L \diff^4 \tau \, 
	\theta_{13} \, \theta_{42} \, 
	D_{12} \, 
	D_{34} \, ,
\end{align*}
such that 
\begin{align*}
F^2 _1 + F^2 _2 = \xi^2 \! \int \limits _0 ^L \diff^4 \tau \, \theta_{42}  \, D_{12} \, D_{34} 
	= \half \xi^2 \! \int \limits _0 ^L \diff^4 \tau \, D_{12} \, D_{34} 
	= \half \left( F^1 _1 \right) ^2  ,
\end{align*}
where we have used the symmetry under $(\tau_1 , \tau_2) \leftrightarrow (\tau_3 , \tau_4)$ to eliminate the ordering of $\tau_2$ and $\tau_4$. Similar relations for the other kinematic factors can be found analogously:
\begin{align}
F^2 _1 + F^2 _2 &= \half \left( F^1 _1 \right) ^2 , &
F^2 _3 + F^2 _4 &= \half \left( F^1 _2 \right) ^2 , &
F^2 _5 &= \left( F^1 _1 \right) ^2 , \\
F^2 _7 + F^2 _8 &= F^1 _1 \, F^1 _2  \, , &
F^2 _9 + F^2 _{10} &= F^1 _1 \, F^1 _2  \, , &  
F^2 _6 &= \left( F^1 _2 \right) ^2  . 
\end{align}
Moreover, we find that the kinematic factors $F^2 _7$ and $F^2 _9$ are related by exchanging $v_1$ and $v_2$. In addition to the above relations, we thus have
\begin{align}
F^2 _9 &= F^2 _7 \, , &
F^2 _{10} &= F^2 _8 \, .
\end{align}
We thus only need to consider the diagrams $F^2 _2$, $F^2 _4$ and the difference $F^2 _7 - F^2 _8$. This finding agrees with the findings from non-Abelian exponentiation. Here, the kinematic factor $F^2_4$ is the color-connected ladder-like diagram appearing in the calculation of the cusp anomalous dimension and $F^2 _7 - F^2 _8$ corresponds to the two-gluon exchange web.

Using the reflection discussed for the one-loop diagrams, we find that $F^2 _2$ and $F^2_4$ are related in the following way: 
\begin{align}
F^2 _4 &= \frac{g^4 \mu^{4 \epsilon}}{\epsilon} 
	\, \xi ^2 \, K ( \phi ) \, , &
F^2 _2 &= \frac{g^4 \mu^{4 \epsilon}}{\epsilon} 
	\, \xi ^2 \, K (\pi - \phi ) \, .
\label{K0def}
\end{align}
For the anomalous dimension, we only need the lowest-order coefficient of 
$K ( \phi ) = K _0 ( \phi ) + \LO ( \epsilon )$, which we compute in appendix 
\ref{app:Kinematic} to find
\begin{align}
K_0 (\phi) = \frac{-1}{2^6 \pi^4  \sin ^2 \phi} \left[
	2\phi \, \mathrm{Im} \left[
	\mathrm{Li}_2 \big( e^{i \phi} \big) + \mathrm{Li}_2 \big( - e^{i \phi} \big)
	\right]
	+ 4 \, \mathrm{Re} \left[
	\mathrm{Li}_3 \big( e^{i \phi} \big) + \mathrm{Li}_3 \big( - e^{i \phi} \big)
	\right]
	- \zeta_3 \right] . 
\label{K0-result}	
\end{align}
The combination $F^2 _7 - F^2 _8$ corresponds to the function $L_0 (\phi)$, 
\begin{align}
F^2 _{7} - F^2 _8 &= \frac{g^4 \mu^{4\epsilon}}{\epsilon} \, \xi ^2
	\left( L_0 (\phi) + \LO ( \epsilon ) \right) , 
\label{L0def}	
\end{align}
which we compute in appendix \ref{app:Kinematic} to find
\begin{align}
L_0 (\phi) &= - \frac{1}{32  \pi ^4 \, \sin^2 \phi} \left[ 
	\phi \, \Img \big( \Li _2 \big( e^{i \phi} \big) \big)
	+ (\pi - \phi) \, \Img \big( \Li _2 \big(- e^{i \phi} \big) \big) \right] . 
\label{L0-result}	
\end{align}
\vspace{-2mm}
\paragraph{Three-vertex and Self-energy Diagrams.} $\mbox{}$\\
Next, we turn to the three-vertex and self-energy diagrams. A cancellation between these diagrams has been observed in the calculation of the cusp anomalous dimension in ref.\ \cite{Makeenko:2006ds}. While the approach presented there has to be adapted slightly to the case of the cross anomalous dimension, the result remains the same. 

The self-energy corrections to the two-point functions have been calculated in ref.\ \cite{Erickson:2000af} using dimensional reduction. They can be accounted for by including a one-loop correction into the scalar propagator, 
\begin{align}
D(x) = g^2 \mu^{2 \epsilon} \, \frac{\Gamma (1- \epsilon )}{4 \pi^{2-\epsilon} } \, 
	\frac{1}{ \left( x^2 \right) ^{1- \epsilon} } 
	- \frac{g^4 \mu ^{4 \epsilon} \, N \, \Gamma ^2 ( 1 - \epsilon)} 
	{32 \, \pi ^{4-2 \epsilon} \, ( 1 - 2 \epsilon ) \, \epsilon} \,
	\frac{1}{ \left( x^2 \right) ^{1- 2 \epsilon} } \, .
\end{align}
Here, the fields are left unrenormalized such that the one-loop correction to the propagator is divergent in 
$\mathrm{D}=4$ dimensions. This is not a problem due to the cancellation with the three-vertex diagrams. For the kinematic factor associated to the self-energy diagram, we then find the expression
\begin{align}
F^2 _{18} =  \frac{N \, \Gamma ^2 ( 1 - \epsilon ) }
	{32 \pi ^4 ( 1 - 2 \epsilon ) \epsilon } 
	\, \left( \pi \mu^2 \right) ^\epsilon 
	\xi
	\int \limits _{-L} ^0 \diff \tau_1 
	\int \limits _0 ^L \diff \tau_2 \, 
	\frac{1}{\big[ \left(x_1 - x_2 \right)^2 \big]^{ 1 - 2 \epsilon} } \, .
	\label{self-energy}
\end{align}
Let us now turn to the three-vertex diagrams. They arise from Wick-contracting the terms
\begin{align*}
& \left \langle \frac{i^3}{3!} \int \diff \tau_1 \, \diff \tau_2 \, \diff \tau_3 \,
	\mathcal{P} \left[ A( \tau_1) A( \tau_2) A( \tau_3 ) \right] 
	_{i i^\prime j j^\prime}
	\left( \frac{1}{g^2 \mu^{2 \epsilon}}
	\int \diff ^\mathrm{D} y f^{a b c} \, \partial ^\mu A^{\nu a} (y) 
	A_\mu ^b (y) A_\nu ^c (y)  \right) 
	\right \rangle \, , \\
& \left \langle \frac{i}{3!} \int \diff \tau_1 \, \diff \tau_2 \, \diff \tau_3 \,
	3 \mathcal{P} \left[ \Phi ( \tau_1) A( \tau_2) \Phi( \tau_3 ) \right]
	_{i i^\prime j j^\prime} 
	\left( \frac{1}{g^2 \mu^{2 \epsilon}}
	\int \diff ^\mathrm{D} y f^{a b c} \, \partial ^\mu \Phi_I ^{a} (y) 
	A_\mu ^b (y) \Phi_I ^c (y) \right) 
	\right \rangle \, ,	
\end{align*}
which gives the term
\begin{align}
\frac{-i}{g^2 \mu^{2 \epsilon}} &\int \diff \tau_1 \, \diff \tau_2 \, \diff \tau_3
	\left( \dot{x}_1 \dot{x}_3 - n_1 n_3 \lvert \dot{x}_1 \rvert \lvert \dot{x}_3 \rvert \right)
	f^{abc} \mathcal{C}(T^a , T^b , T^c ) _{i i^\prime j j^\prime} \nn \\
\quad \cdot &\int \diff ^\mathrm{D} y \, \Big( \dot{x}_2 \frac{\partial}{\partial y} \, D(x_1 - y) \Big) 
	D(x_2 - y ) D(x_3 -y ) = \nn \\
= \frac{i}{g^2 \mu^{2 \epsilon}} 
	&\int \diff \tau_1 \, \diff \tau_2 \, \diff \tau_3
	\left( \dot{x}_1 \dot{x}_3 - n_1 n_3 \lvert \dot{x}_1 \rvert \lvert \dot{x}_3 \rvert \right)
	\dot{x}_2 \frac{\partial}{\partial x_1} G(x_1 , x_2 , x_3)
	f^{abc} \mathcal{C}(T^a , T^b , T^c ) _{i i^\prime j j^\prime} \, .	
	\label{Three-Gluon-Contribution}	
\end{align}
Here, $G(x_1 , x_2 , x_3 ) = \int \diff ^\mathrm{D} y \,   D(x_1 - y) D(x_2 - y ) D(x_3 -y )$ and $\mathcal{C}$ orders 
$T^a , T^b , T^c$ according to the ordering of $(\tau_1 , \tau_2 , \tau_3)$ and assigns the appropriate indices. 

Note that we have exhausted our freedom to interchange the integration variables $(\tau_1, \tau_2, \tau_3)$ in carrying out the contractions. For any given diagram we must hence consider all combinations of $(\tau_1, \tau_2, \tau_3)$ being on any of the edges. We see immediately that the integral vanishes for diagrams with $\tau_1$ and $\tau_3$ on the same line. In the case of $\tau_2$ and $\tau_3$ being on the same line, we have $\dot{x}_2 = \dot{x}_3$ and $n_2 = n_3$, such that upon exchanging $\tau_2$ and $\tau_3$ only the color factor $f^{abc} \mathcal{C}(T^a , T^b , T^c ) _{i i^\prime j j^\prime}$ changes its sign. These diagrams thus vanish as well and we need only consider those contributions with $(\tau_1 , \tau_2 )$ on one line and $\tau_3$ on the other. For these diagrams, we still need to take the two possible orderings of $(\tau_1 , \tau_2 )$ into account. A different order changes the sign of the color factor. It is convenient to include this into the kinematic factor rather than calculating the diagrams separately and piece them together only after summing over diagrams weighted with different color factors. Symbolically we then compute as follows: 
\begin{align*}
C \left( \, \raisebox{-7mm}{\includegraphics[height=10ex]{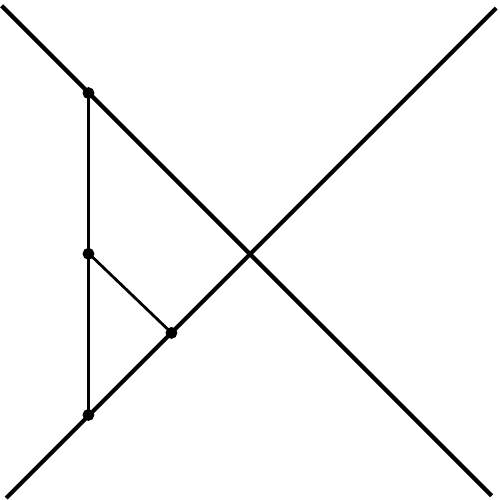}} \, \right)
 	& =  f^{abc}  \, \raisebox{-7mm}{\includegraphics[height=10ex]{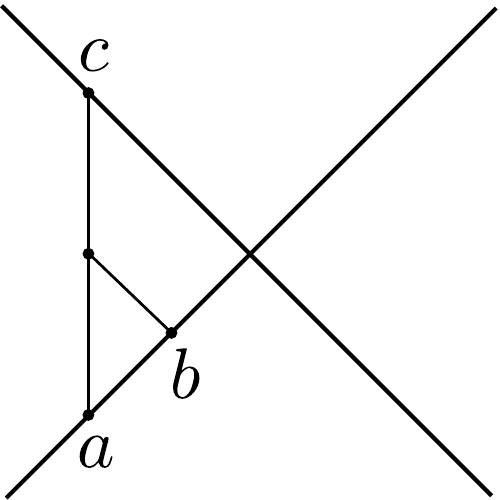}} \, , \qquad 
F \left( \, \raisebox{-7mm}{\includegraphics[height=10ex]{Figures/colorkinplain.pdf}} \, \right)
	= F \left( \, \raisebox{-7mm}{\includegraphics[height=10ex]{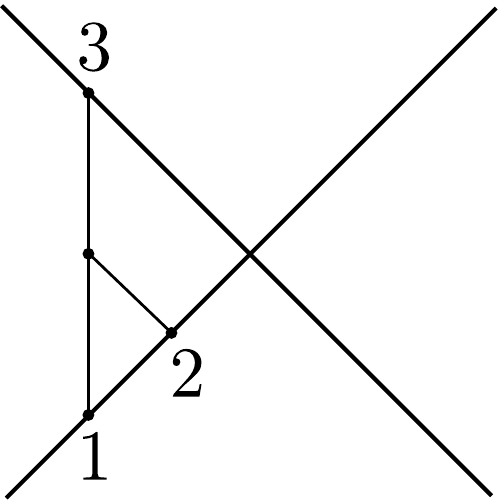}} \, \right) - 
	F \left( \, \raisebox{-7mm}{\includegraphics[height=10ex]{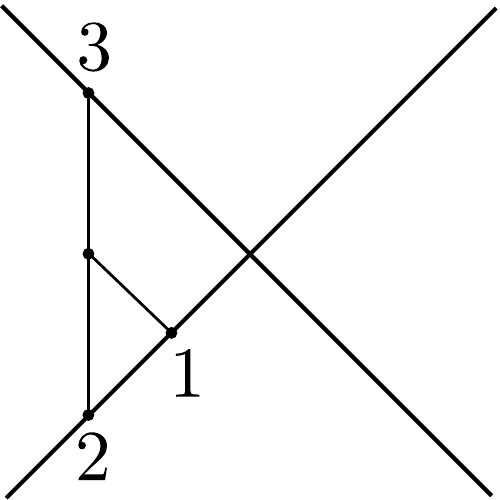}} \, \right)  .
\end{align*}
In the case of $\tau_1$ and $\tau_2$ being on the same line, we may identify 
$\dot{x}_2 \, \partial/ \partial x_1 = \partial / \partial \tau_1$,
since $\dot{x}_2 = \dot{x}_1$. For the kinematic factor $F^2 _{14}$ we thus have
\begin{align}
F^2 _{14} & = \frac{i \xi }{g^2 \mu^{2 \epsilon}} 
	\int \limits _{-L} ^0 \diff \tau_1 \, \diff \tau_2 \,
	\varepsilon ( \tau_2 - \tau_1 ) 
	\int \limits _0 ^L \diff \tau_3 \,
	\frac{\partial}{\partial \tau_1} \, G(x_1 , x_2 , x_3) \nn \\
& = \frac{ 2 i \xi }{g^2 \mu^{2 \epsilon}} 
	\int \limits _{-L} ^0 \diff \tau_1 \, 
	\int \limits _0 ^L \diff \tau_3 \,
	G(x_1 , x_1 , x_3) 
	- \frac{ i \xi }{g^2 \mu^{2 \epsilon}} 
	\int \limits _{-L} ^0 \diff \tau_2 \,  
	\int \limits _0 ^L \diff \tau_3 \,
	G( 0 , x_2 , x_3) 
\label{FStep}
\end{align}
Here, we have left out the contribution from the upper boundary $\tau_1 = L$, since it is not relevant for the pole term.

\begin{figure}[t]
\centering
\includegraphics[width=85mm]{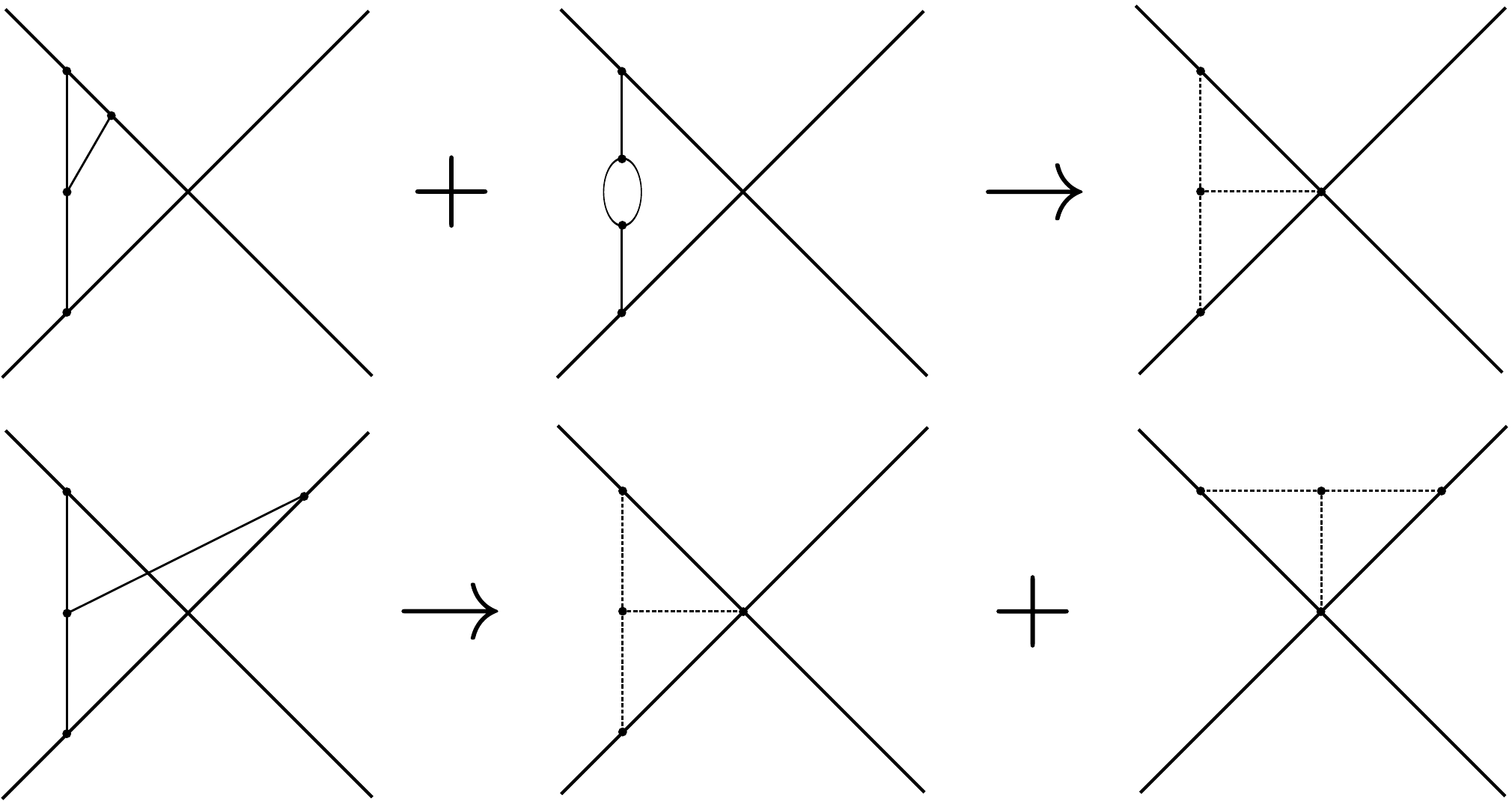}
\caption{Symbolic display of the relations for the three-vertex and self-energy diagrams.}
\label{fig:FrozenOnCusp}
\end{figure}

By introducing Feynman parameters, we can rewrite the three-point correlator $G(x_1,x_2,x_3)$ as
\begin{align}
G(x_1 , x_2 , x_3 ) &= \left( \frac{ g^2 \mu^{2 \epsilon} \Gamma ( 1 - \epsilon ) }
	{4 \pi^{2 - \epsilon} } \right) ^3 
	\int \diff ^\mathrm{D} y \, 
	\frac{1}{ \left[ (x_1 -y )^2 (x_2 -y )^2 (x_3 -y )^2 \right]^{1 - \epsilon} } \nn \\
	&= g^6 \mu^{6 \epsilon} \, \frac{\Gamma(1- 2 \epsilon) }{4^3 \pi^{4 - 2 \epsilon} } 
	\int \limits _0 ^1 \diff \alpha \, \diff \beta \, \diff \gamma \, 
	\frac{\delta (1 - \alpha - \beta - \gamma ) (\alpha \beta \gamma)^{- \epsilon}  }
	{\left[ \alpha \beta x_{12} ^2 + \beta \gamma x_{23} ^2 + \alpha \gamma x_{13} ^2 \right] ^{1 - 2 \epsilon} } \, .
\label{3ptdef}	
\end{align}
Here, we have integrated out $y$ according to the formula
\begin{align*}
\int \diff ^\mathrm{D} y \frac{1}{\left[ y^2 + \Delta \right]^\kappa } = \frac{\Gamma \big( \kappa - \ft{\mathrm{D}}{2} \big) }{\Gamma ( \kappa ) } \, \frac{\pi ^{\mathrm{D}/2}}{\Delta^{\kappa - \mathrm{D}/2}} \, .
\end{align*}
For the first term in equation \eqref{FStep}, we then note 
\begin{align*}
G(x_1 , x_1 , x_3 ) &=  \frac{\Gamma(1- 2 \epsilon) g^6 \mu^{6 \epsilon} }
	{4^3 \pi^{4 - 2 \epsilon} \, (x_{13} ^2 )^{1 - 2 \epsilon} } \, 
	\int \limits _0 ^1  \diff \alpha \, \diff \beta \, \diff \gamma
	\frac{\delta ( 1 - \alpha - \beta - \gamma ) ( \alpha \beta \gamma ) ^{- \epsilon} }
	{[ \gamma  ( 1 - \gamma ) ] ^{1 - 2 \epsilon}} \\
&= \frac{g^6 \mu^{6 \epsilon} \,\Gamma ^2 ( 1 - \epsilon) }
	{4^3 \pi ^{4 - 2 \epsilon} ( 1 - 2 \epsilon ) \epsilon }  
	\, \frac{1}{(x_{13} ^2 )^{1 - 2 \epsilon}} \, .
\end{align*}
The comparison with \eqref{self-energy} then shows that
\begin{align}
F^2 _{14} &= \frac{i}{N} \, F^2 _{18}  
	- \frac{ i \xi }{g^2 \mu^{2 \epsilon}} 
	\int \limits _{-L} ^0 \diff \tau_2 \,  
	\int \limits _0 ^L \diff \tau_3 \,
	G( 0 , x_2 , x_3)
	= \frac{i}{N} \, F^2 _{18} 
	+ \frac{g^4 \mu^{4 \epsilon}}{\epsilon} \, \xi \, F( \phi) \, ,
\label{F0def}	
\end{align}
and after summing over all diagrams with the respective combinatorial and color factors, the kinematic factor $F^2 _{18}$ drops out. The three-vertex contribution thus reduces to the case of one scalar being frozen to the intersection point, see also figure \ref{fig:FrozenOnCusp}. It is described by the lowest-order coefficient of 
$F(\phi) = F_0 (\phi) + \LO(\epsilon) $, which we compute in appendix \ref{app:Kinematic} to find
\begin{align}
F_0 (\phi) = - \frac{2 i \phi \left( \pi^2 - \phi ^2 \right)}
	{( 4 \pi ) ^4 \, 3 \sin \phi}   \, .
\end{align}
The other three-vertex diagrams can be discussed analogously. This gives the following relations: 
\begin{align}
F^2 _{11} &= F^2 _{12} 
	= \frac{i}{N} \, F^2 _{17}  
	+ \frac{g^4 \mu^{4 \epsilon}}{\epsilon} \, \xi \, F( \pi - \phi) \, , & \nn \\
\label{F0rel}	
F^2 _{13} &= F^2 _{14} 
	= \frac{i}{N} \, F^2 _{18}  
	+ \frac{g^4 \mu^{4 \epsilon}}{\epsilon} \, \xi \, F( \phi) \, , & \\	
F^2 _{15} &= F^2 _{16} 
	= - \frac{g^4 \mu^{4 \epsilon}}{\epsilon} 
		\, \xi \, ( F( \phi) + F( \pi - \phi) )  \, , & \nn
\end{align}
For the three-vertex contributions $F^2 _{15}$ and $F^2 _{16}$ connecting three of the semi-infinite lines, we find two boundary terms corresponding to the two orderings of $\tau_1$ and $\tau_2$. After freezing one scalar to the intersection point, the contributions only involve two of the semi-infinite lines and can be expressed by the function $F(\phi)$ which also appears in the cusp anomalous dimension, see figure \ref{fig:FrozenOnCusp}. We have thus found that the three-vertex contributions to the cross anomalous dimension can be related to the respective contributions to the cusp anomalous dimension. 

\subsection{Relation to the Cusp Anomalous Dimension}
It is clear that some of the above kinematic factors appear also in the calculation of the cusp anomalous 
dimension at the two-loop level. Specifically, these are the factors $F^2_3$, $F^2_4$, $F^2_{13}$, $F^2_{14}$ 
and $F^2_{18}$ and hence the functions $F_0 (\phi)$ and $K_0 (\phi)$. Combining these with the appropriate color 
factors and using the relations derived for the kinematic factors above, we find the cusp anomalous dimension 
\begin{align}
\Gamma _{\mathrm{cusp}} ^{(2)} (\phi , \rho ) 
	= \frac{N^2 - 1}{N^2} \, \gacu ^{(2)} 	(\phi , \rho )
	= \frac{N^2 - 1}{N^2} \left( -\xi^2 \, K_0 (\phi) + 2 i \xi \, F_0 (\phi) \right) . 
\end{align} 


\subsection{The Two-loop Result}
In terms of the color and kinematic factors, the correlator $\mathcal{W}_a$ at the two-loop level is given by 
\begin{align}
\mathcal{W}_a  ^{\,(2)} = \sum \limits _m \ft{1}{N^2} \, c_m \, F^2 _m \, C^2 _{a, m} \, .
\end{align}
Here, the $c_n$ denote combinatorial factors, for which we have
\begin{align*}
c_5 = c_6 = 1 \, , \qquad c_n = 2 \quad \text{for} \quad n \neq 5 , 6 \, .
\end{align*}
The color factors $C^2 _{a, m}$ are collected in appendix \ref{app:Color}. Using the relations between the kinematic factors discussed above, it is then straightforward to express the cross anomalous dimension in terms of the functions 
$K_0 (\phi)$, $F_0 (\phi)$ and $L_0 (\phi)$ by using equations \eqref{K0def}, \eqref{L0def} and \eqref{F0def}. Due to the presence of the commutator term at the two-loop level, we also need the finite coefficients of the kinematic factors at the one-loop level,
\begin{align}
F^1 _1 &= \frac{g^2 \mu^{2\epsilon}}{\epsilon} ( \cos \phi - \cos \rho )
	\left( I_0 (\pi - \phi) + \epsilon \,  I_1 (\pi - \phi) + \LO ( \epsilon^2 ) \right) , \\
F^1 _2 &= \frac{g^2 \mu^{2\epsilon}}{\epsilon} ( \cos \phi - \cos \rho )
	\left( I_0 ( \phi) + \epsilon \,  I_1 ( \phi) + \LO ( \epsilon^2 ) \right)  . 
\end{align}
The relevant combination $ C_0(\phi) = I_0 (\phi) I_1(\pi - \phi) - I_0 (\pi - \phi) I_1(\phi)$ 
is computed in appendix \ref{app:Kinematic}, where we find that
\begin{align}
C_0(\phi)	= - \frac{1}{32 \pi^4 \sin^2 \phi} 
	\Big[ (\phi + \pi) \Img \big( \Li_2 \big( e^{i \phi} \big) \big)
	- (\phi - 2 \pi) \Img \big( \Li_2 \big(- e^{i \phi} \big) \big) \Big] . 
\label{C0-result}	
\end{align}
From the renormalization group equation \eqref{rge:gacrt}, we then find the two-loop coefficient of the cross anomalous dimension to be given by
\begin{align}
\Gamma _{\mathrm{cross}} ^{(2)} (\phi,\rho) &= 
	\frac{ 2 \widetilde{\Gamma} _{\mathrm{cusp}} ^{(2)} (\phi,\rho)} {N^2}
	\begin{pmatrix}
	0 & 0 \\ - 1 & N^2
	\end{pmatrix} 
	+ \frac{2 \Gamma _{A} ^{(2)} (\phi,\rho)} {N^2} 
	\begin{pmatrix}
	-1 & N^2 \\ 0 & 0 
	\end{pmatrix}  
	+ \frac{2 \Gamma _{B} ^{(2)} (\phi,\rho)} {N^2} 
	\begin{pmatrix}
	0 & 0 \\ - 1 & 1
	\end{pmatrix} .
\end{align}
Here, we have introduced the abbreviations 
\begin{align}
\Gamma _{A} ^{(2)} (\phi,\rho) &=  \xi^2 \left( 
	K_0 (\pi - \phi) - K_0 (\phi) + C_0 (\phi) - L_0 (\phi) \right) , \\
\Gamma _{B} ^{(2)} (\phi,\rho) &= - \xi^2 
	\left( 	K_0 (\pi - \phi) - K_0 (\phi) - C_0 (\phi) + L_0 (\phi) \right) 
	- 4 i \xi \left( F_0 (\phi) + F_0 (\pi - \phi) \right) .	
\end{align}
It is noteworthy that by virtue of the relations \eqref{F0rel} all three-vertex and 
self-energy contributions to the first line of the cross anomalous dimension cancel out. 
Remarkably, also the contribution of the ladder diagrams%
\footnote{This behavior has also been observed for the cross anomalous dimension in QCD \cite{Korchemskaya:1994qp}.} 
to the first line of the cross anomalous dimension vanishes, which can be seen by inserting the results 
\eqref{K0-result}, \eqref{L0-result} and \eqref{C0-result} to find 
\begin{align}
\Gamma _{A} ^{(2)} (\phi,\rho) = 0 \, . 
\end{align}
Using the same identity, we can rewrite the contribution $\Gamma _{B} ^{(2)} (\phi,\rho)$ to get 
\begin{align}
\Gamma _{B} ^{(2)} (\phi,\rho) &= - 2 \left( 
	- \xi^2 K_0 (\phi) + 2 i \xi F_0 (\phi) 
	+ \xi^2 K_0 (\pi - \phi) + 2 i \xi F_0 (\pi - \phi) \right) \nn \\
	&= -2 \gacu ^{(2)} (\phi,\rho)
	   +2 \gacu ^{(2)} (\pi - \phi,\pi - \rho) . 
	= -2 \gacu ^{(2)} (\phi,\rho)
	   +2 \gacur ^{(2)} (\phi, \rho) .   
\end{align}
The cross anomalous dimension can thus indeed be expressed in terms of the cusp anomalous 
dimension also at the two-loop level, 
\begin{align}
\Gamma _{\mathrm{cross}} ^{(2)} 
	&= \begin{pmatrix}
	0 & 0 \\
	0 & 2 \gacu ^{(2)}
	\end{pmatrix}	 
	- \frac{2}{N^2} 
	\begin{pmatrix}
	0 & 0 \\ 
	2 \gacur ^{(2)} - \gacu ^{(2)}  & 2 \gacu ^{(2)} - 2 \gacur ^{(2)}
	\end{pmatrix} . 
\label{two-loop-result}	
\end{align}
A consistency check%
\footnote{I would like to thank Gregory Korchemsky for pointing out this option.}
of the above result can be performed by analytically continuing to a 
Min\-kows\-kian cusp angle $i \gamma$ and considering the limit $\gamma \to \infty$. It is known \cite{Korchemskaya:1996je}
that in this limit the eigenvalues $\Gamma_{\pm}$ of the cross anomalous dimension matrix scale as
\begin{align}
\Gamma _{+} &= \gamma \, \Gamma (\lambda) + \mathrm{O} ( \gamma ^0) \, , & 
\Gamma _{-} &= \mathrm{O} ( \gamma ^{-1}) \, .
\label{scaling}
\end{align}
Here, $\Gamma (\lambda)$ is related to the respective coefficient of the cusp anomalous dimension in the limit 
$\gamma \to \infty$, i.e.\ when the cusp becomes lightlike. 
For the analytic continuation, note that the $i 0$-prescription for the position space propagator implies that 
$\gamma$ has a small negative imaginary part, such that the branch cuts of the polylogarithms appearing for terms 
such as $\Li_2 (e^\gamma)$ are approached from below.  

The scaling \eqref{scaling} can be reduced to the requirement that the trace of the cross anomalous dimension matrix 
scales as $\gamma$ whereas its determinant scales as $\gamma^0$. This is easily checked for the result 
\eqref{two-loop-result}, since $\gacu$ scales linearly in $\gamma$ and the determinant vanishes. For the one-loop 
result, one notes that the determinant can be expressed in terms of the combination $(\gacu - \gacur)$, which scales 
as $\gamma^0$ for $\gamma \to \infty$. 
Note that, even though the passing of our above result was easily established, the check does put a strong constraint on the 
result since the contributions of the individual diagrams at the two-loop order scale as $\gamma^3$, such that the correct 
scaling behavior involves a number of cancellations, e.g.\ between ladder and three-vertex diagrams as it is the case for 
the cusp anomalous dimension. 

Let us now turn to the two-loop calculation for the contour shown in figure \ref{fig:mixing3}. As already at the 
one-loop level, we reconsider the diagrams given in figure \ref{fig:two-loop} and find that the kinematic factors 
$\Ft ^2 _i$ can be related to the kinematic factors $F ^2 _i$ we have calculated above. 
Consider for example the first diagram for which we find the contribution 
\begin{align}
\Ft ^2 _1 &= (-\xi)^2 \, \frac{\Gamma(1-\epsilon)^2}{16 \pi ^{4-2\epsilon}} g^4 \mu ^{4 \epsilon} \, 
	\int \limits _{-L} ^0 \diff \tau_2 \, \diff \tau_4  
	\int \limits _0    ^L \diff \tau_1 \, \diff \tau_3 \, 
	\frac{\theta(\tau_3 - \tau_1) \theta(\tau_2 - \tau_4) }
		{\big( \left( v_2 \tau_1 + v_1 \tau_2 \right)^2 
			\left( v_2 \tau_3 + v_1 \tau_4 \right)^2  \big)^{1-\epsilon}}	
	\nn \\
	&= \xi ^2 \, \frac{\Gamma(1-\epsilon)^2}{16 \pi ^{4-2\epsilon}} g^4 \mu ^{4 \epsilon} \, 
	\int \limits _0 ^L \diff^4 \tau \, 
	\frac{\theta(\tau_3 - \tau_1) \theta(\tau_4 - \tau_2) }
		{\big( \left( v_2 \tau_1 - v_1 \tau_2 \right)^2 
			\left( v_2 \tau_3 - v_1 \tau_4 \right)^2  \big)^{1-\epsilon}}	= F ^2 _1 \, .
\end{align}
All ladder diagrams can be discussed similarly. For the three-vertex diagrams such as 15 and 16 in figure 
\ref{fig:two-loop} we should take into account that opposite lines have opposite orientations, $\dot{x}_2 = - \dot{x}_1$, 
such that we have $\dot{x}_2 \cdot \partial_{x_1} = - \partial_{\tau_1}$ in contrast to calculation of the kinematic factors 
$F^2 _i$ and the diagrams 11 - 14. We thus find
\begin{align}
\Ft ^2 _{15} = - \frac{i \xi}{g^2 \mu ^{2 \epsilon} } 
	\int \limits _0 ^L \diff^3 \tau \, 
	\partial_{\tau_1} \left[ G ( -v_2 \tau_1 , v_2 \tau_2 , v_1 \tau_3 )
		+ G ( v_2 \tau_1 , - v_2 \tau_2 , v_1 \tau_3 ) \right] 
	= F ^2 _{15} \, .
\end{align}
In summary, all kinematic factors can be related to their respective counterparts up to a sign, which is given by
\begin{align}
\Ft ^2 _i = \begin{cases}  
	F ^2 _i \quad \text{for} \quad 
		i \in \left \lbrace 1, \ldots, 6 , 13, 14 , 15, 18 \right \rbrace , \\
	- F ^2 _i \quad \text{for} \quad 
		i \in \left \lbrace 7, \ldots, 12 , 16, 17 \right \rbrace .
	\end{cases}
\end{align}
The associated color factors are collected in appendix \ref{app:Color}. Using the same steps as described above for the 
cross anomalous dimension $\Gacr$, we then find
\begin{align}
\Gacrt ^{(2)} = 2 \begin{pmatrix}
	\gacur ^{(2)} & 2 \gacu ^{(2)} - \gacur ^{(2)} \\
	0 & \gacu ^{(2)}
	\end{pmatrix}
	- \frac{2}{N^2} \begin{pmatrix}
	2 \gacu ^{(2)} & 0 \\
	\gacu ^{(2)} - 2 \gacur ^{(2)} & 2 \gacur ^{(2)}
	\end{pmatrix} .
\end{align}
Remarkably, the cross anomalous dimension $\Gacrt$ can also be expressed in terms of the cusp anomalous dimension 
at the two-loop order as we have observed at strong coupling. 

\section{Conclusion and Outlook}
\label{sec:outlook}

In this paper, we have determined the cross anomalous dimension for Maldacena--Wilson loops in 
$\mathcal{N} \!=4$ SYM theory at strong coupling and up to the two-loop level at weak coupling. 
The strong-coupling discussion showed that the cross anomalous dimension displays Gross--Ooguri 
phase transitions in certain regions of the parameter space. 
A further goal of this calculation was to investigate whether a relation between the cross and cusp 
anomalous dimension can be established, which indeed turned out to be possible. 

A further check of this structure would require a to carry out a three-loop calculation. At this level, 
one can benefit from the understanding of non-Abelian exponentiation \cite{Gardi:2010rn,Mitov:2010rp} 
and modern methods for the calculation of kinematic factors \cite{ArkaniHamed:2012nw,Lipstein:2012vs}. 
Moreover, a subset of diagrams will again be immediately related to the cusp anomalous dimension, 
which is currently known to four loops \cite{Correa:2012nk,Henn:2013wfa}. 

It would be very interesting to see whether it is possible --- as it is the case for the cusp anomalous 
dimension in $\mathcal{N} \!=4$ SYM theory ---  to calculate the cross anomalous dimension or limits thereof 
based on localization \cite{Correa:2012at} or integrability \cite{Drukker:2012de,Correa:2012hh}, see also refs. 
\cite{Gromov:2015dfa,Gromov:2016rrp,Cavaglia:2018lxi} for recent developments on the cusp anomalous dimension. 
The cross anomalous dimension provides a richer structure than the cusp anomalous dimension (e.g.\ the 
strong-coupling phase transition) and the availability of exact methods could lead to interesting checks of the AdS/CFT 
correspondence. 

\section*{Acknowledgements}
I would like to thank Konstantin Zarembo for pointing me to this project and for his guidance in its early phase.  
Moreover I would like to thank Johannes Broedel, Johannes Henn, Dennis M{\"u}ller, Jan Plefka, Amit Sever and in 
particular Christoph Sieg for interesting discussions. I am also indebted to Konstantin Zarembo and in particular 
to Gregory Korchemsky for their very valuable comments on the draft.
Additionally, I would like to thank Shota Komatsu and Simone Giombi for pointing 
out typos in an earlier version. 

This work has been supported by the SFB 647 \textit{``Space-Time-Matter.Analytic and Geometric Structures''} and the Research 
Training Group GK 1504 \textit{``Mass, Spectrum, Symmetry''} of the German Research Foundation 
as well as the grant no.\ 615203 from the European Research Council under the FP7 and by the 
Swiss National Science Foundation through the NCCR SwissMAP. 

\newpage
\appendix

\section{Color Factors}
\label{app:Color}
In this appendix, we provide the color factors arising from the contraction of the $\mathrm{SU(N)}$-generators $T^a$ for the calculation of the cross anomalous dimension to two loops.
For the structure constants, we note the convention
$\left[ T^a , T^b \right] = i \, f^{a b c} \, T^c$. The generators are normalized in such a way that $ 2 \tr \left(T^a T^b \right) = \delta ^{a b}$ and we note the following basic 
$\mathrm{SU(N)}$-identities:
\begin{equation}
\begin{alignedat}{2}
T^a _{i j} T^a _{k l} &= \half \left( \delta_{i l} \delta_{k j} - \ft{1}{N} \delta_{i j} \delta_{k l} \right) \, , & 
\qquad \qquad
T^a T^a  &= \ft{N^2-1}{2N} \, \mathbf{1} \, , \\
T^a T^b T^a &= - \ft{1}{2N} T^b \, ,  & \qquad \qquad
f^{a b c} f^{a b d} &= N \delta^{cd} \, .
\end{alignedat}
\end{equation}
These identities are sufficient to calculate the color factors at the one- and two-loop level. All color factors can be expressed as linear combinations of the two basic color structures 
\begin{align}
\one &= \delta _{i i^\prime} \, \delta _{j j^\prime} 
\, , &
\two  &= \delta _{i j^\prime} \, \delta _{j i^\prime} 
\, .
\end{align}
At the one-loop level, we note the color factors
\begin{align}
C_{1,1} ^1  &= \half \two - \ft{1}{2N} \one \, , & \qquad  
C_{1,2} ^1  &= C_{1,1} ^1 \, , & \qquad 
C_{1,3} ^1  &= C_{1,1} ^1 \, , & \qquad 
C_{1,4} ^1  &= C_{1,1} ^1 \, , \\
C_{2,1} ^1  &= \half \one - \ft{1}{2N} \two \, , & \qquad  
C_{2,2} ^1  &= \ft{N^2 - 1}{2N} \two , & \qquad 
C_{2,3} ^1  &= C_{2,1} ^1 \, , & \qquad 
C_{2,4} ^1  &= C_{2,2} ^1 \, .
\end{align}
At the two-loop level, we first consider the color factors for $\mathcal{W}_1$. 
For the ladder-like diagrams we note
\begin{equation}
\begin{alignedat}{4}
C_{1,1}^2 &= \ft{N^2 +1}{4 N^2} \one - \ft{1}{2N} \two \, , & \qquad 
C_{1,2}^2 &= \ft{N^2 -2}{4 N} \two + \ft{1}{4N^2} \one \, , & \qquad
C_{1,3}^2 &= C_{1,2}^2 \, , & \qquad 
C_{1,4}^2 &= C_{1,1}^2 \, , \\
C_{1,5}^2 &= C_{1,1}^2  \, , & \qquad
C_{1,6}^2 &= C_{1,2}^2  \, ,  & \qquad
C_{1,7}^2 &= C_{1,1}^2  \, , & \qquad 
C_{1,8}^2 &= C_{1,2}^2  \, , \\
C_{1,9}^2 &= C_{1,1}^2  \, , & \qquad
C_{1,10}^2 &= C_{1,2}^2  \, .  & \qquad
\end{alignedat}
\end{equation}
For the three-vertex diagrams we have
\begin{align}
C_{1,11} ^2 = C_{1,12} ^2 = \ldots = C_{1,16} ^2 = \ft{i N}{4} \two - \ft{i}{4} \one  ,
\end{align}
and for the self-energy diagrams we get
\begin{align}
C_{1,17} ^2 = C_{1,1} ^1 \, , \qquad  C_{1,18} ^2 = C_{1,2} ^1 \, .
\end{align}
Note that the relation $C_{1,11} ^2 = \ft{i N}{2} C_{1,17} ^2$ between the color factors of the three-vertex and self-energy diagrams is essential for the cancellations between the three-vertex and self-energy diagrams. 

For the operator $\mathcal{W}_2$ we find the following color factors for the ladder-like diagrams: 
\begin{equation}
\begin{alignedat}{3}
C_{2,1}^2 &= \ft{N^2 +1}{4 N^2} \two - \ft{1}{2N} \one \, , & \quad 
C_{2,2}^2 &= \ft{N^2 -2}{4 N} \one + \ft{1}{4N^2} \two \, , & \quad
C_{2,3}^2 &= \ft{(N^2-1 )^2}{4 N^2} \two  \, , \\
C_{2,4}^2 &= - \ft{N^2-1}{4 N^2} \two \, , & \quad
C_{2,5}^2 &= C_{2,1}^2 \, , & \quad
C_{2,6}^2 &= C_{2,3}^2  \, ,  \\
C_{2,7}^2 &= \ft{N^2-1}{4N} \one - \ft{N^2-1}{4N^2} \two  \, , & \quad  
C_{2,8}^2 &= - \ft{1}{4N} \one + \ft{1}{4N^2} \two \, , & \quad
C_{2,9}^2 &= C_{2,7}^2  \, ,  & \quad
C_{2,10}^2 &= C_{2,8}^2 \, .
\end{alignedat}
\end{equation}
For the three-vertex diagrams we have
\begin{equation}
\begin{alignedat}{4}
C_{2,11}^2 &= \ft{i N}{4} \one - \ft{i}{4} \two \, , & \qquad \quad
C_{2,12}^2 &= C_{2,11}^2  \, , & \qquad \quad
C_{2,13}^2 &=  \ft{i (N^2-1)}{4} \two \, , & \\
C_{2,14}^2 &= C_{2,13}^2\, , & \qquad \quad
C_{2,15}^2 &= - C_{2,11}^2  \, , & \qquad \quad
C_{2,16}^2 &= - C_{2,11}^2   \, ,  & 
\end{alignedat}
\end{equation}
and for the self-energy diagrams we get
\begin{align}
C_{2,17} ^2 &= C_{2,1} ^1 \, , &  C_{2,18} ^2 &= C_{2,2} ^1 \, .
\end{align}
The color factors for the contour shown in figure \ref{fig:mixing3} at the one-loop level have been 
stated in the main text. At the two-loop level, we find the following color factors 
for the loop $\widetilde{\mathcal{W}}_1$: 
\begin{align}
\Ct_{1,1}^2 &= \ft{(N^2-1)^2}{4 N^2} \one  \, , & 
\Ct_{1,2}^2 &= - \ft{N^2-1}{4 N^2} \one \, , & 
\Ct_{1,5}^2 &= \Ct_{1,1}^2  \, , & 
\Ct_{1,6}^2 &= \Ct_{1,3}^2 \, , \nn \\
\Ct_{1,3}^2 &= \ft{N^2-2}{4N} \two + \ft{1}{4N^2} \one \, , & 
\Ct_{1,4}^2 &= \ft{N^2+1}{4N^2} \one - \ft{1}{2N} \two \, ,  & 
\Ct_{1,9}^2 &= \Ct_{1,7}^2  \, , & 
\Ct_{1,10}^2 &= \Ct_{1,8}^2  \, , \\ 
\Ct_{1,7}^2 &= - \ft{1}{4N} \two + \ft{1}{4N^2} \one  \, , &
\Ct_{1,8}^2 &=  \ft{N^2 - 1}{4N} \two - \ft{N^2 - 1}{4N^2} \one   \, . & \nn
\end{align}
For the three-vertex diagrams we have
\begin{equation}
\begin{alignedat}{4}
\Ct_{1,11}^2 &= \ft{i (N^2-1)}{4} \one \, , & \qquad \quad
\Ct_{1,12}^2 &= \Ct_{1,11}^2  \, , & \qquad \quad
\Ct_{1,13}^2 &=  \ft{i N}{4} \two - \ft{i}{4} \one \, , & \\
\Ct_{1,14}^2 &= \Ct_{1,13}^2\, , & \qquad \quad
\Ct_{1,15}^2 &= - \Ct_{1,13}^2  \, , & \qquad \quad
\Ct_{1,16}^2 &= \Ct_{1,13}^2   \, .  & 
\end{alignedat}
\end{equation}
and for the self-energy diagrams we get
\begin{align}
\Ct_{1,17} ^2 &= \Ct_{1,1} ^1 \, , &  \Ct_{1,18} ^2 &= \Ct_{1,2} ^1 \, .
\end{align}
For the loop $\widetilde{\mathcal{W}}_2$, the relation to the untilded case is 
simpler: The color factors are the same, 
\begin{align}
\Ct_{2,i} ^2 = C_{2,i} ^2 ,
\end{align}
except for the following cases: 
\begin{align}
\Ct_{2,1} ^2 &= C_{2,2} ^2 \, , &
\Ct_{2,2} ^2 &= C_{2,1} ^2 \, , &
\Ct_{2,5} ^2 &= C_{2,2} ^2 \, , &
\Ct_{2,15} ^2 &= - C_{2,15} ^2 \, .
\end{align}

\section{Kinematic Factors}
\label{app:Kinematic}

In this appendix, we calculate the relevant integrals for the cross anomalous dimension up to the two-loop level. While powerful methods for the evaluation of integrals of the kind encountered here exist \cite{Davydychev:1992xr,ArkaniHamed:2012nw,Lipstein:2012vs}, we shall take a pedestrian approach below, taking inspiration from ref.\ \cite{Korchemsky:1987wg}. The results of the integrals calculated below are typically given in terms of polylogarithms. The polylogarithm functions can be defined recursively by
\begin{align}
\frac{\diff }{\diff z} \, \Li_n(z) &= 
	\frac{1}{z} \, \Li_{n-1} (z) \, , &
	\Li_1 (z) &= - \ln ( 1- z ) \, . 
\label{polylog}	
\end{align}
Here, we encounter in particular the dilogarithm function
\begin{align}
\Li_2 (z) = - \int \limits _0 ^z \diff u \, 
	\frac{\ln (1-u) }{u} \, ,
\label{dilogdef}	
\end{align}
which satisfies the functional identities 
\begin{align}
\Li_2 \Big( \frac{1}{z} \Big) = -\Li_2 (z) 
	- \frac{\pi^2}{6} - \ln ^2 (-z) \, , \qquad
	\Li_2 ( 1 - z ) = -\Li_2 (z) 
	+ \frac{\pi^2}{6} - \ln (z) \ln (1-z) \, .
\end{align}

\subsection{Cusp Anomalous Dimension}
We begin by studying the integrals which appear also in the calculation of the cusp anomalous dimension. Explicit results for these were derived in ref.\ \cite{Drukker:2011za}, partly based on the results of ref.\ \cite{Davydychev:1992xr}.
We present an independent, elementary derivation below. 

\vspace*{-2mm}
\paragraph{Three-vertex Diagrams.} $\mbox{}$\\
We compute the function $F_0 ( \phi )$, which encodes the contribution of the three-vertex diagram to the cusp anomalous dimension. Starting from equation \eqref{F0def}, we have
\begin{align}
F (  \phi ) = - \frac{i \epsilon}{g^6 u^{6 \epsilon} } \, 
	\int \limits _{-L} ^0 \diff \tau_2 \,
	\int \limits _0 ^L \diff \tau_3 \,
	G ( 0 , x_2 , x_3 ) \, .
\end{align}
After plugging in the expression \eqref{3ptdef} for $G ( 0 , x_2 , x_3 )$, substituting $\tau_2 = - \kappa z$, $\tau_3 = \kappa \zb$ as in the one-loop calculation and carrying out the $\kappa$-integral, we find 
\begin{align}
F _0 (  \phi ) & = - \frac{i }{( 4 \pi ) ^4} \, 
	\int \limits _0 ^1 \diff z \, \diff \alpha \, \diff \beta \, \diff \gamma \, 
	\frac{\delta(1 - \alpha - \beta - \gamma) }
	{\alpha \beta z^2 + \beta \gamma ( z^2 + \zb^2 + 2 z \zb \cos \phi ) + \alpha \gamma \zb^2 } \nn \\
	&= - \frac{i }{( 4 \pi ) ^4} \,
	 \int \limits _0 ^1 \diff z \, \diff \gamma
	\int \limits _0 ^{\gamma} \diff \beta \,
	\frac{1}{\beta \bar{\beta} z^2 + 2 \beta \bar{\gamma} z \zb \cos \phi + \gamma \bar{\gamma} \zb^2} \, .
\end{align}
Here, we have integrated out $\alpha$ and substituted $\gamma \rightarrow \bar\gamma$ to reach a more convenient form. The integral can be further facilitated by substituting 
\begin{align*}
r = \frac{\zb}{z} \, , \qquad 
u = \frac{\beta}{\gamma} \, , \qquad 
v = \frac{\bar\beta}{\bar\gamma} \, , 
\end{align*}
to reach the form
\begin{align*}
F_0 ( \phi ) & =  - \frac{i }{( 4 \pi ) ^4} \, 
	\int \limits _0 ^{\infty} \diff r
	\int \limits _0 ^{1} \diff u
	\int \limits _1 ^{\infty} \diff v
	\frac{1}{(v-u)(v u+ 2 u r \cos \phi + r^2) } \\
	& = - \frac{i }{( 4 \pi ) ^4} \, 
	\int \limits _0 ^{\infty} \diff r
	\int \limits _0 ^{1} \diff u
	\int \limits _1 ^{\infty} \diff v
	\frac{1}{(v-u)(v + u (r^2 + 2 r \cos \phi) ) } \, . 
\end{align*}
After carrying out the $v$-Integration, we have 
\begin{align*}
F_0 ( \phi ) &= \frac{1}{( 4 \pi ) ^4 \, 2 \sin \phi}
	\int \limits _0 ^{\infty} \diff r
	\int \limits _0 ^{1} \diff u \,
	\left( \frac{1}{r + e^{i \phi}}	- \frac{1}{r + e^{-i \phi}}	\right) 
	\frac{\ln \left( 1 + u (r^2 + 2r \cos \phi) \right) - \ln \left( 1 - u \right)  } 
	{u} \\
	& =: - \frac{i f_0 (\phi)}{( 4 \pi ) ^4 \, 2 \sin \phi} \, .
\end{align*}
Rather than calculating $f_0(\phi)$ directly, we consider its derivative. Noting that 
\begin{align*}
\partial_\phi \left( \frac{1}{r + e^{i \phi}}	- \frac{1}{r + e^{-i \phi}}	\right) =
i \, \partial_r \left( \frac{e^{i\phi}}{r + e^{i \phi}} + \frac{e^{-i\phi}}{r + e^{-i \phi}} \right) ,
\end{align*}
and after integrating by parts, one finds (after some cancellations)  
\begin{align*}
f_0^\prime (\phi) & = 4 \cos \phi \, \int \limits _0 ^\infty \diff r \, 
	\int \limits _0 ^{1} \diff u \,
	\frac{1}{1+ u (r^2+ 2 r \cos \phi)}
	- 2 \int \limits _0 ^{1} \diff u \,
	\frac{\ln (1-u)}{u}  \\
	& = \frac{\pi^2}{3} + 4 \cos \phi \, \int \limits _0 ^\infty \diff r \, 
	\frac{\ln \left(r^2 + 2 r \cos \phi +1 \right) }{r (r + 2 \cos \phi )}
	\, .
\end{align*}
The anti-derivative of the above integrand can be constructed by decomposing into partial fractions and factoring the logarithm. In order to take the limits $r \to 0$ and $r \to \infty$, one may rely on the known asymptotic expansions of polylogarithms. Then, one finds that  
\begin{align*}
f_0^\prime (\phi) = \frac{4 \pi^2}{3} - 4 \phi^2 
\qquad \Rightarrow \qquad
f_0(\phi) = \frac{4}{3} \phi \left( \pi^2 - \phi ^2 \right) ,
\end{align*}
where we have used $f_0(0)=0$. We have thus found that
\begin{align}
F_0 (\phi) = - \frac{2 i \phi \left( \pi^2 - \phi ^2 \right)}
	{( 4 \pi ) ^4 \, 3 \sin \phi}   \, .
\end{align}

\paragraph{Crossed Propagators Diagram.} $\mbox{}$\\
The ladder-like contribution to the cusp anomalous dimension is encoded in the diagram $F^2_4$, for which we note the expression
\begin{align}
F^2_4 &= \frac{g^4}{16 \pi^4} \left( \pi \mu^2 L^2 \right)^{2 \epsilon} 
	\xi ^2
	\int \limits _0 ^1 
	\frac{\diff^4 \tau \,\theta( \tau_4 - \tau_3 ) \, \theta( \tau_1 - \tau_2 ) 
	\, \Gamma^2(1- \epsilon) }
	{\left[ \left(\tau_1 ^2 + \tau_3 ^2 + 2 \tau_1 \tau_3 \cos \phi \right)
	\left(\tau_2 ^2 + \tau_4 ^2 + 2 \tau_2 \tau_4 \cos \phi \right)	  
	  \right] ^{1-\epsilon}}  \nn \\
	  &= \frac{g^4 \mu ^{4\epsilon}}{\epsilon} 
	\xi^2 \, \left(  K_{0} ( \phi )  + \LO ( \epsilon ) \right)  \, .
\end{align}
in order to calculate $K_{0} (\phi)$, we substitute
\begin{align}
\tau_1 &= \kappa z \,, & 
\tau_2 &= x \kappa z \,, & 
\tau_3 &= \kappa \zb \,, & 
\tau_4 &= y \kappa \zb \, . 
\end{align} 
Here, we have $x,y,z \in [0,1]$ and since we are again only interested in the divergent piece, taking $\kappa \in [0,1]$ is sufficient. Then, we arrive at
\begin{align}
K_0 (\phi) &= \frac{1}{2^6 \pi ^4} 
	\int \limits _0 ^1 \diff x \, \diff y \, \diff z 
	\frac{z \zb}
	{ \left( z^2 + y^2 \zb^2 + 2 y z \zb \cos \phi \right)
	  \left( x^2 z^2 + \zb^2 + 2 x z \zb \cos \phi \right) } \nn \\
	  &= - \frac{1}{2^8 \pi^4 \sin ^2 \phi }
	  \int \limits _0 ^1 \frac{\diff z}{z \zb} \,
	  \ln \left( \frac{\zb + z e^{i \phi}}{\zb + z e^{-i \phi}} \right) \,
	  \ln \left( \frac{z + \zb e^{i \phi}}{z + \zb e^{-i \phi}} \right)  .
\end{align}
Here, we have restricted $\phi$ to the domain $\phi < \pi/2$, such that we may work with the standard form of the complex logarithm, having the branch cut along the negative real axis. The result for $\phi \geq \pi/2$ can be deduced from analytic continuation. In the above expression, we substitute 
\begin{align}
e ^{2 i \psi} = \frac{z + \zb e^{i \phi}}{z + \zb e^{-i \phi}} 
\end{align}
to reach the form
\begin{align*}
K_{0} (\phi) &= \frac{1}{2^6 \pi^4  \sin ^2 \phi} 
	\int \limits _0 ^\phi \diff \psi \,
	\psi ( \phi - \psi ) \left( \cot(\psi) + \cot (\phi - \psi) \right) 
	= \frac{1}{2^5 \pi^4  \sin ^2 \phi} 
	\int \limits _0 ^\phi \diff \psi \,
	\psi ( \phi - \psi ) \cot(\psi) \, .
\end{align*}
In order to compute the integral 
\begin{align}
k_0 ( \phi ) = \int \limits _0 ^\phi \diff \psi \,
	\psi ( \phi - \psi ) \cot(\psi) \, , 
\end{align}
it is convenient to consider $k_0^\prime ( \phi)$, for which one finds the simpler integral 
\begin{align*}
k_0^\prime ( \phi) = \int \limits _0 ^\phi \diff \psi \, \psi  \cot(\psi) 
	= -i \int \limits _1 ^{e^{i\phi}} \diff u \,
	\frac{ (u^2 + 1) \ln{(u)} }
	{u (u^2 - 1)} \, .
\end{align*}
In the second expression, we have substituted $u = e^{i \psi }$ to reach a standard form of logarithmic integrals. By decomposing into partial fractions, one arrives at   
\begin{align}
k_0^\prime ( \phi) = \phi \, \mathrm{Re} \left[ 
	\ln \big( 1 + e^{i\phi} \big) + \ln \big( 1 - e^{i\phi} \big) \right]
	+ \mathrm{Im} \left[
	\mathrm{Li}_2 \big( e^{i \phi} \big) +
	\mathrm{Li}_2 \big( - e^{i \phi} \big)  \right]  .
\end{align}
We may integrate the above expression by noting that due to equation \eqref{polylog}, we have 
\begin{align*}
\partial_\phi \, \mathrm{Li}_n \big( e^{i\phi} \big) =  i \, \mathrm{Li}_{n-1} \big( e^{i\phi} \big) \, , 
\quad \text{where} \quad 
\mathrm{Li}_1 \big( e^{i\phi} \big) = - \ln \big( 1 - e^{i\phi} \big) \, .
\end{align*}
Enforcing the boundary condition $k_0(0) = 0$, we then reach 
\begin{align}
k_0(\phi) = - \phi \, \mathrm{Im} \left[ 
	\mathrm{Li}_2 \big( e^{i \phi} \big) + \mathrm{Li}_2 \big( - e^{i \phi} \big)
	\right]
	- 2 \, \mathrm{Re} \left[
	\mathrm{Li}_3 \big( e^{i \phi} \big) + \mathrm{Li}_3 \big( - e^{i \phi} \big)
	\right] 
	+ \frac{\zeta_3}{2} \, , 
\end{align}
such that we arrive at
\begin{align}
K_0 (\phi) = \frac{-1}{2^6 \pi^4  \sin ^2 \phi} \left[
	2\phi \, \mathrm{Im} \left[
	\mathrm{Li}_2 \big( e^{i \phi} \big) + \mathrm{Li}_2 \big( - e^{i \phi} \big)
	\right]
	+ 4 \, \mathrm{Re} \left[
	\mathrm{Li}_3 \big( e^{i \phi} \big) + \mathrm{Li}_3 \big( - e^{i \phi} \big)
	\right]
	- \zeta_3 \right] . 
\end{align}
The result agrees with the one given in ref.\ \cite{Drukker:2011za} after making use of 
functional identities for poly\-loga\-rithms. 
\subsection{Cross Anomalous Dimension}
\paragraph{Commutator Term.} $\mbox{}$\\
The commutator term appearing in the cross anomalous dimension at the two-loop level involves the finite contributions to the diagrams $F^1 _1$ and $F^1 _2$. We recall the expressions
\begin{align*}
F^1 _2 &= -g^2 \frac{\Gamma(1-\epsilon)}{4 \pi ^2} 
	\left( \pi \mu^2 L^2 \right) ^\epsilon 
	\int \limits _0 ^1 \diff \tau_1 \diff \tau_2 \,
	\frac{\left( \cos \phi - \cos \rho \right) }
	{\left( \tau_1 ^2 + \tau_2 ^2 + 2 \tau_1 \tau_2 \cos \phi \right) 
	^{1- \epsilon} } 
	= \frac{g^2 \mu^{2 \epsilon} }{\epsilon}
	\xi \, I(\phi)  , 
\end{align*}
and similarly
\begin{align*}
F^1 _1 = \frac{g^2 \mu^{2 \epsilon} }{\epsilon}
	\xi \, 	I(\pi - \phi) .
\end{align*}
In order to compute $I(\phi) = I_0 (\phi) + \epsilon I_1 ( \phi) + \LO ( \epsilon^2 )$, we substitute $\tau_1 = \kappa z$, $\tau_2 = \kappa \zb$ as before. As we are interested in the finite part of the integral, taking $\kappa \in [0,1]$ is not sufficient. Rather, we have $z \in [0,1]$ and $\kappa \in [0, f(z)]$ with 
\begin{align*}
f(z) = \begin{cases}
1/ \zb \quad &\text{for} \quad z \in [0, \half ] \\
1/ z   \quad &\text{for} \quad z \in [\half , 1 ] \, .
\end{cases} 
\end{align*}
Then we have
\begin{align*}
I(\phi) &= - \frac{(\pi L^2)^\epsilon \, \epsilon \, \Gamma ( 1 - \epsilon)  }{4 \pi^2} 
	\int \limits _0 ^1 \diff z
	\frac{1}{\left( z^2 + \zb^2 + 2 z \zb \cos \phi \right)^{1- \epsilon}  } 
	\int \limits _0 ^{f(z)} \diff \kappa \, \kappa ^{-1+2 \epsilon} \\
	&=  - \frac{1 + \alpha \epsilon}{8 \pi^2} 
	\int \limits _0 ^1 
	\frac{\diff z }{a(z) a(\zb) } 
	- \frac{\epsilon}{8 \pi^2}
	\int \limits _0 ^1 \diff z \,
	\frac{\ln \left( a(z) a(\zb)   \right) }
	{ a(z) a(\zb)} 
	+ \frac{\epsilon}{2 \pi^2}
	\int \limits _0 ^{1/2} \diff z \,
	\frac{\ln \left( \zb  \right) }
	{a(z) a(\zb)} + \LO(\epsilon^2) ,
\end{align*}
where we abbreviated $a(z) := z e^{i \phi/2} + \zb e^{-i \phi/2}$ and the precise value 
of the expansion coefficient $\alpha$ is irrelevant for us. 
The integrals appearing in $I_{1} (\phi)$ can be reduced to integrals of the form 
\eqref{dilogdef} by decomposing into partial fractions. Recalling also our result 
\eqref{I0-result}, 
\begin{align*}
I_0 (\phi) = - \frac{\phi}{8 \pi^2 \sin \phi} \, ,  
\end{align*}
we obtain 
\begin{align}
I_{1} (\phi) &= 
	\frac{-1}{4 \pi^2 \sin \phi} \left[
	\phi  \ln (2 \sin ( \phi /2 ) \sin \phi ) 
	+ \Img \big( 
	2 \, \Li_2 \big( e^{i \phi} \big) 
	+ 2 \, \Li_2 \big( \half \big( 1 - e^{i \phi} \big) \big) 	
	+  \Li_2 \big( - e^{i \phi} \big) \big) + \ft{\alpha}{2} \, \phi
	\right] \nn \\
	&= -\frac{1}{4 \pi^2 \sin \phi} \left[
	\phi  \ln (2 \sin \phi ) 
	+ \Img \big( 
	\Li_2 \big( e^{i \phi} \big) 
	+ 2 \, \Li_2 \big( - e^{i \phi} \big) \big) 
	+ \ft{\alpha}{2} \, \phi \right] . 
\end{align}
In the last step, we have made use of the identity
\begin{align*}
2 \, \Img \big( \Li_2 \big( \half \big( 1 - e^{i \phi} \big) \big) \big) 
	+ \phi \ln ( \sin ( \phi /2 ) ) 
	= \Img \big( \Li_2 \big( - e^{i \phi} \big) - \Li_2 \big( e^{i \phi} \big) \big) , 
\end{align*}
which the reader should have no problem to verify by taking a $\phi$-derivative. 
For the combination $C_0(\phi) = I_0 (\phi) I_1(\pi - \phi) - I_0 (\pi - \phi) I_1(\phi)$ 
appearing in the commutator term, we then find
\begin{align}
C_0(\phi)	= - \frac{1}{32 \pi^4 \sin^2 \phi} 
	\Big[ (\phi + \pi) \Img \big( \Li_2 \big( e^{i \phi} \big) \big)
	- (\phi - 2 \pi) \Img \big( \Li_2 \big(- e^{i \phi} \big) \big) \Big] . 
\label{result:Commutator-Term}	
\end{align}

\paragraph{Two-gluon Exchange Web.} $\mbox{}$\\
We calculate the difference between the kinematic factors $F^2 _7$ and $F^2_8$, which corresponds to the two-gluon exchange web. For $F^2_7$, we have the expression
\begin{align}
F^2 _7 	&= \frac{g^4 \, \Gamma(1-\epsilon)^2}{16 \pi ^4} \left( \pi \mu^2 L^2\right)^{2 \epsilon} \xi ^2
	\int \limits _0 ^1 
	\frac{ \diff ^4 \tau \, \theta ( \tau_2 - \tau_3 )  }
	{\left[  \left(\tau_1^2 + \tau_2^2 - 2 \tau_1 \tau_2 \cos \phi \right)     
	\left(\tau_3^2 + \tau_4^2 + 2 \tau_3 \tau_4 \cos \phi \right)
	\right]^{1-\epsilon} } \, .
\end{align}
Similar to the one-loop case, we substitute
\begin{align*}
\tau_2 = \kappa z \, , \qquad
\tau_3 = x \kappa z \, , \qquad
\tau_1 = y \kappa \bar{z} \, , \qquad
\tau_4 = x \bar{y} \kappa \bar{z} \, .
\end{align*}
For the coordinate ranges we note $x,y,z \in [0 , 1 ]$. Since $F^2 _7$ has a double pole and we are interested in the single pole term, we need to determine the upper boundary for $\kappa$, which is given by 
\begin{align*}
f(x,y,z) = \kappa \left[ \mathrm{max} 
\lbrace \tau_1 , \tau_2 , \tau_3 , \tau_4 \rbrace \right] ^{-1} \, .
\end{align*}
Let us however first discuss the integral, for which we find
\begin{align*}
F^2 _7  =  \frac{g^4 \, \Gamma(1-\epsilon)^2}{16 \pi ^4} 
	\left( \pi \mu^2 L^2\right)^{2 \epsilon} 
	\xi ^2 
	\int \limits _0 ^1 \diff y \, \diff z 
	\frac{z \bar{z}}
	{ R(y,z) ^{1 - \epsilon} } 
	 \int \limits _0 ^1 \diff x \, x^{-1+2 \epsilon} 
	\int \limits _0 ^{f(x,y,z)} \diff \kappa \, \kappa^{-1+4 \epsilon} \, .
\end{align*}
where we abbreviated
\begin{align*}
R(y,z) =  
	\left( z^2 + y^2 \bar{z}^2 -2 y z \bar{z} \cos \phi \right)
	\left( z^2 + \bar{y}^2 \bar{z}^2 + 2 \bar{y} z \bar{z} \cos \phi \right) .	
\end{align*}
We observe that the poles stem from the $x$ and $\kappa$ integrations. In determining the upper boundary for $\kappa$, we need to discriminate between different cases. In the case $\tau_4 > \tau_1$ for example, we have $x > y/\bar{y}$, such that the pole term of the $x$-Integration disappears. In this case, the upper boundary for $\kappa$ is irrelevant for the single pole term and we may set it equal to 1. In this way, we find
\begin{align}
f(x,y,z) \simeq \begin{cases}
	1 \quad &\text{if} \quad x > y/\bar{y} \\
	1/z \quad &\text{if} \quad x < y/\bar{y} \, , y < z/\bar{z} \\
	1/(y\zb) \quad &\text{if} \quad x < y/\bar{y} \, , y > z/\bar{z} \, .
\end{cases}
\label{lambda_boundaries}
\end{align} 
We can discuss the factor $F^2_8$ similarly. In the expression 
\begin{align}
F^2 _8
	&= \frac{g^4 \,\Gamma(1-\epsilon)^2}{16 \pi ^4} \left( \pi \mu^2 L^2\right)^{2 \epsilon} \xi^2 
	\int \limits _0 ^1 
	\frac{ \diff ^4 \tau \, \theta ( \tau_3 - \tau_2 )  }
	{\left[  \left(\tau_1^2 + \tau_2^2 - 2 \tau_1 \tau_2 \cos \phi \right)     
	\left(\tau_3^2 + \tau_4^2 + 2 \tau_3 \tau_4 \cos \phi \right)
	\right]^{1-\epsilon} } \, ,
\end{align}
we substitute
\begin{align*}
\tau_2 = x \kappa z \, , \qquad
\tau_3 = \kappa z \, , \qquad
\tau_1 = x y \kappa \bar{z} \, , \qquad
\tau_4 = \bar{y} \kappa \bar{z} \, .
\end{align*}
The upper boundary for $\kappa$ is then found as $f(x , \bar{y} , z)$, such that we have 
\begin{align*}
F^2 _8  =  \frac{g^4 \, \Gamma(1-\epsilon)^2}{16 \pi ^4} 
	\left( \pi \mu^2 L^2\right)^{2 \epsilon} 
	\xi ^2 
	\int \limits _0 ^1 \diff y \, \diff z 
	\frac{z \bar{z}}
	{ R(\bar{y},z) ^{1 - \epsilon} } 
	 \int \limits _0 ^1 \diff x \, x^{-1+2 \epsilon} 
	\int \limits _0 ^{f(x,y,z)} \diff \kappa \, \kappa^{-1+4 \epsilon} 
	+ \LO (\epsilon^0)  . 
\end{align*}
We can then combine the two kinematic factors to find
\begin{align*}
F^2 _7 - F^2 _8 &=  \frac{g^4 \mu^{4\epsilon}}{\epsilon} \xi ^2
	\left( L_0 (\phi) + \LO ( \epsilon ) \right) , \\
L_0 (\phi) &= \frac{\epsilon}{16  \pi ^4} \,
	\int \limits _0 ^1 \diff x \, \diff y \, \diff z 
	\int \limits _1 ^{f(x,y,z)} \diff \kappa \, 
	x^{-1 + 2 \epsilon} \kappa^{-1 + 4 \epsilon} 
	\left( \frac{z \zb }{R(y,z) ^{1 - \epsilon} }
	- \frac{z \zb }{R(\bar{y},z) ^{1 - \epsilon} } \right) .
\end{align*}
Here, we have already used that the integral vanishes due to the antisymmetry under $y \leftrightarrow \bar{y}$ when we restrict the integration domain of $\kappa$ to the interval from 0 to 1. After plugging in the boundary values \eqref{lambda_boundaries}, we are left with
\begin{align}
L_0 (\phi) &= \frac{1}{32  \pi ^4}
	\int \limits _0 ^{1/2}  \diff z 
	\int \limits _{z/ \zb} ^1 \diff y \, \ln \Big( y \frac{\zb}{z} \Big)
	\left( \frac{ z \zb }{R(\bar{y},z) }
	- \frac{z \zb }{R(y,z) } \right) .  
\end{align}  
Now, we substitute
\begin{align*}
r = \frac{\zb}{z} \, , \qquad
x = y \, \frac{\zb}{z} \, , 
\end{align*}
which leaves us with
\begin{align}
L_0 (\phi) &= \frac{1}{32  \pi ^4} 
	\int \limits _1 ^\infty \! \diff r \!
	\int \limits _1 ^r 
	\frac{\diff x}{(1+r)^4}  
	\left( \frac{\ln (x) }{R\big( \ft{r-x}{r}  , \ft{1}{1+ r} \big) }
	- \frac{ \ln (x) }{R \big( \ft{x}{r}  , \ft{1}{1+ r} \big) } \right)  
	= \frac{1}{32  \pi ^4} \left( l_0 (\pi - \phi) - l_0 (\phi) \right) . 
\end{align}
Here, we introduced the function $l_0 (\phi)$, which is given by
\begin{align*}
l_0 (\phi) &= \int \limits _1 ^\infty \diff x \,
	\frac{\ln ( x ) }{ \big( x - e^{i \phi} \big) \big( x - e^{-i \phi} \big)  } 
	\int \limits _x ^\infty \diff r \, 
	\frac{1}{\big( r - x + e^{i \phi} \big) \big( r - x + e^{-i \phi} \big)   } \\
	& = \int \limits _1 ^\infty \diff x \,
	\frac{\ln ( x ) }{ \big( x - e^{i \phi} \big) \big( x - e^{-i \phi} \big)  } 
	\int \limits _0 ^\infty \diff r \, 
	\frac{1}{\big( r + e^{i \phi} \big) \big( r + e^{-i \phi} \big)   } 
	= \frac{\phi}{ \sin ^2 \phi } \,  
	\Img \big( \Li _2 \big( e^{i \phi} \big) \big) .
\end{align*}
We have thus found
\begin{align}
L_0 (\phi) &= - \frac{1}{32  \pi ^4 \, \sin^2 \phi} \left[ 
	\phi \, \Img \big( \Li _2 \big( e^{i \phi} \big) \big)
	+ (\pi - \phi) \, \Img \big( \Li _2 \big(- e^{i \phi} \big) \big) \right] . 
\end{align}
When comparing these results with similar calculations in the literature, one should 
note that the single diagrams typically depend on the infrared regulator that is being used.

\newpage

\bibliographystyle{nb}
\bibliography{biblio}

\end{document}